\documentclass[a4paper,11pt]{article}
\pdfoutput=1 

\usepackage{jheppub} 
\usepackage{amssymb} 
\usepackage{amsmath}
\usepackage{mathtools}
\usepackage{amsfonts}    
\usepackage{dsfont}
\usepackage{pdfpages}
\usepackage{verbatim}
\usepackage{linearA}
\usepackage{tensor}
\usepackage{mathrsfs}
\usepackage[mathscr]{euscript}
\usepackage{tikz}
\usepackage{braket}

\usetikzlibrary{calc,decorations.markings}
\usetikzlibrary{decorations.pathmorphing}
\usetikzlibrary{shapes,backgrounds}
\usetikzlibrary{fadings}
\usetikzlibrary{cd}
\tikzset{
    partial ellipse/.style args={#1:#2:#3}{
        insert path={+ (#1:#3) arc (#1:#2:#3)}
    }
}

\allowdisplaybreaks
\usepackage{mathrsfs}
\usepackage[mathscr]{euscript}

\numberwithin{equation}{section}

\def\be{\begin{equation}}
\def\ee{\end{equation}}

\title{\boldmath Higher genus correlators for tensionless $\rm{AdS}_3$ strings}

\author{Bob Knighton}

\affiliation{Institut f\"ur Theoretische Physik, ETH Z\"urich \\
\hspace*{0.3cm} Wolfgang-Pauli-Stra{\ss}e 27, 8093 Z\"urich, Switzerland}

\emailAdd{robejr@ethz.ch}

\abstract{It was recently shown in \cite{Dei:2020} that tree-level correlation functions in tensionless string theory on $\rm{AdS}_3\times\rm{S}^3\times\mathbb{T}^4$ match the expected form of correlation functions in the symmetric orbifold CFT on $\mathbb{T}^4$ in the large $N$ limit. This analysis utilized the free-field realization of the $\mathfrak{psu}(1,1|2)_1$ Wess-Zumino-Witten model, along with a surprising identity directly relating these correlation functions to a branched covering of the boundary of $\rm{AdS}_3$. In particular, this identity implied the unusual feature that the string theory correlators localize to points in the moduli space for which the worldsheet covers the boundary of $\rm{AdS}_3$ with specified branching near the insertion points. In this work we generalize this analysis past the tree-level approximation, demonstrating its validity to higher genus worldsheets, and in turn providing strong evidence for this incarnation of the $\rm{AdS}/\rm{CFT}$ correspondence at all orders in perturbation theory.}

\begin{document}

\maketitle

\section{Introduction}

Recently, progress has been made in providing a derivation of a special case of the $\rm{AdS}/\rm{CFT}$ correspondence \cite{Maldacena_1999} relating type IIB string theory on $\rm{AdS}_3\times\rm{S}^3\times\mathbb{T}^4$ with $k=1$ units of pure NS-NS flux to the symmetric orbifold $\text{Sym}^N(\mathbb{T}^4)$ in the large $N$ limit. It has been shown that the full tree-level spectra of both theories match \cite{Gaberdiel_2018,Eberhardt:2018}. It was also known that the correlation functions of the symmetric orbifold theory obey nontrivial Ward identities for strings on $\rm{AdS}_3$ both at tree level \cite{Eberhardt:2019} and at higher genus \cite{Eberhardt:2020} in the RNS formalism. However, it was not shown that the symmetric orbifold correlators were the unique solutions to the Ward identities. Other recent approaches toward the realization of this duality have been explored in \cite{Hikida_2020,Sfondrini:2020ovj,Gaberdiel:2020ycd}.

\begin{figure}[!ht]
\centering
\begin{tikzpicture}
\draw[smooth] (0,1) to[out=30,in=150] (2,1) to[out=-30,in=210] (3,1) to[out=30,in=150] (5,1) to[out=-30,in=30] (5,-1) to[out=210,in=-30] (3,-1) to[out=150,in=30] (2,-1) to[out=210,in=-30] (0,-1) to[out=150,in=-150] (0,1);
\draw[smooth] (0.4,0.1) .. controls (0.8,-0.25) and (1.2,-0.25) .. (1.6,0.1);
\draw[smooth] (0.5,0) .. controls (0.8,0.2) and (1.2,0.2) .. (1.5,0);
\draw[smooth] (3.4,0.1) .. controls (3.8,-0.25) and (4.2,-0.25) .. (4.6,0.1);
\draw[smooth] (3.5,0) .. controls (3.8,0.2) and (4.2,0.2) .. (4.5,0);
\fill (0,0) circle (0.05);
\node[above] at (0,0) {$z_1$};
\fill (2,0.5) circle (0.05);
\node[below] at (2,0.5) {$z_2$};
\fill (3,-0.5) circle (0.05);
\node[above] at (3,-0.5) {$z_3$};
\fill (5,0) circle (0.05);
\node[above] at (5,0) {$z_4$};
\draw[thick, -latex] (6,0) -- (8.5,0);
\node[above] at (7.25,0) {$\Gamma$};
\draw (11,0) circle (2);
\draw (11,0) [partial ellipse=180:360:2 and 0.5];
\draw[dashed] (11,0) [partial ellipse=0:180:2 and 0.5];
\fill (11,1) circle (0.05);
\node[above] at (11,1) {$x_1$};
\fill (12,-1) circle (0.05);
\node[above] at (12,-1) {$x_2$};
\fill (10,0) circle (0.05);
\node[above] at (10,0) {$x_3$};
\fill (10.5,-1.5) circle (0.05);
\node[above] at (10.5,-1.5) {$x_4$};
\end{tikzpicture}
\caption{A cartoon representation of a holomorphic covering map $\Gamma:\Sigma\to\rm{S}^2$ mapping the string theory worldsheet to the dual CFT sphere. Correlation functions in both tensionless string theory and the symmetric product CFT are determined by the data encoded in $\Gamma$.}
\label{fig:covering-map}
\end{figure}
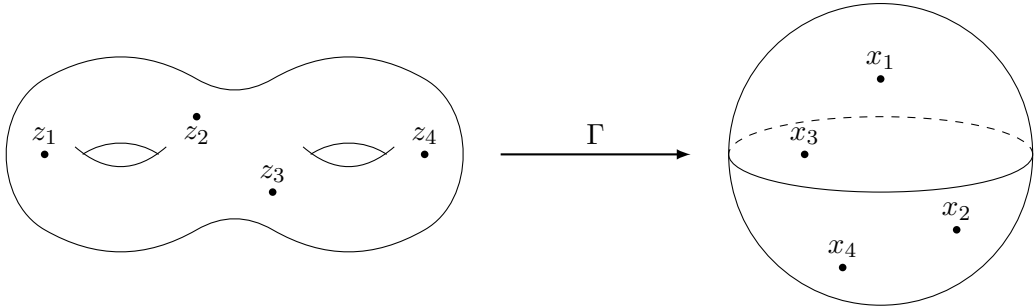

The central property that relates the correlation functions of tensionless string theory to those of the symmetric orbifold is that both are related to a certain meromorphic function $\Gamma:\Sigma\to\rm{S}^2$ which covers $\rm{S}^2$ by the worldsheet. From the perspective of the symmetric orbifold, $\Gamma$ plays the role of a conformal transformation under which the pullback of the twisted fields inserted into correlators become single-valued \cite{Hamidi:1986vh,Lunin_2001,Lunin_2002}, as shown in Figure \ref{fig:covering-map}. On the string theory side, this holomorphic function is interpreted as a map from the string worldsheet to the boundary of $\rm{AdS}_3$, for which the incoming and outgoing string states asymptotically wind $w_i$ times at the points $x_i$ on the boundary. The interpretation of this incarnation of the $\rm{AdS}_3/\rm{CFT}_2$ duality is, then, that the string worldsheet is exactly the covering space used in the calculation of symmetric orbifold correlation functions, as was originally proposed in \cite{Pakman_2009}. As such, one would expect the worldsheet correlators to localize to those points in the moduli space for which the worldsheet is such a covering space. That such a solution was permitted by the worldsheet Ward identities in the RNS formalism was the primary result of \cite{Eberhardt:2019,Eberhardt:2020}.

In \cite{Dei:2020}, another approach for constraining correlation functions for the tensionless ($k=1$) string was employed, similar in spirit to that of \cite{Eberhardt:2019,Eberhardt:2020}. This approach was based on the fact that, at $k=1$, the hybrid formalism of Berkovits, Vafa and Witten \cite{Berkovits_1999} simplifies into (a quotient of) a free field theory on the worldsheet, consisting of four spin-$1/2$ Bosons and four spin-$1/2$ Fermions. The Ward identities, when expressed in these variables, turn out to be enough to essentially fix the form of the correlation functions of highest-weight states in the tensionless string at tree level. The crux of this analysis was a striking formula, reminiscent of a twistorial identity, relating the worldsheet correlation functions to the covering map $\Gamma:\Sigma\to\rm{S}^2$. This identity directly realizes the localization of the worldsheet correlation functions, and allows the interpretation of the string worldsheet as the covering space in Figure \ref{fig:covering-map}, at the level of string perturbation theory. The main advantage of the analysis of \cite{Dei:2020} was that this localizing solution was found to be the only one consistent with the Ward identities in the free field description.

The goal of this paper is to extend the results of \cite{Dei:2020} to worldsheets of higher genus. On the string theory side, this corresponds to considering correlation functions at higher-order in perturbation theory. On the side of the dual CFT, this corresponds to considering correlation functions at higher order in the $1/N$ expansion. We find that an identity analogous to the main result of \cite{Dei:2020} is satisfied on higher genus worldsheets, and thus that the conclusions of \cite{Dei:2020} holds at every order in perturbation theory. This represents highly nontrivial evidence that tensionless string theory on $\rm{AdS}_3\times\rm{S}^3\times\mathbb{T}^4$ is exactly dual to the symmetric product CFT on $\mathbb{T}^4$.

This paper is organized as follows. Section \ref{sec:Worldsheet} introduces the relevant details of the free field construction of tensionless string theory on $\rm{AdS}_3\times\rm{S}^3\times\mathbb{T}^4$. In particular, we formulate string theory on this spacetime in terms of free Bosons generating the algebra $\mathfrak{sl}(2,\mathbb{R})_1\oplus\widehat{\mathfrak{u}(1)}$. We introduce the vertex operators corresponding to spectrally-flowed highest-weight states of this algebra, and define their correlation functions after taking into account picture-changing. In Section \ref{sec:Incidence-relation}, we prove an identity relating worldsheet correlation functions to the covering map used to construct correlation functions in the symmetric product CFT, which generalizes the identity found in \cite{Dei:2020} to higher genus worldsheets. This identity is essentially what allows us to show that the worldsheet correlation functions localize on the moduli space, and thus interpret the string worldsheet as the covering surface appearing in the construction of the symmetric orbifold correlation functions \cite{Hamidi:1986vh,Lunin_2001}. In Section \ref{sec:Ward-to-covering}, we show that the constraints on correlation functions arising from taking various OPE limits (reminiscent of Ward identities) are algebraically identical to the constraints used to construct covering a covering map $\Gamma:\Sigma\to\rm{S}^2$, thus providing a second method of directly relating the string correlation functions to the covering space. We then conclude the main text with a discussion in Section \ref{sec:discussion}. Additionally, we include a brief introduction to the relevant results from the theory of Riemann surfaces in Appendix \ref{sec:Riemann-Surfaces} which are used heavily throughout the main text.

\section{The worldsheet theory}\label{sec:Worldsheet}

In the RNS formulation of strings on $\rm{AdS}_3\times\rm{S}^3\times\mathbb{T}^4$ with $k$ unites of pure NS-NS flux, one identifies the $\rm{AdS}_3$ factor with the $\mathcal{N}=1$ supersymmetric Wess-Zumino-Witten model on the Lie group $\text{SL}(2,\mathbb{R})$ at level $k$ \cite{Maldacena-Ooguri-1,Maldacena-Ooguri-2,Maldacena-Ooguri-3}, while identifying the $\rm{S}^3$ factor as a similar WZW model on $\text{SU}(2)$. Symbolically, the tensionless string can be formulated in terms of the worldsheet WZW model

\begin{equation}
\mathfrak{sl}(2,\mathbb{R})^{(1)}_k\oplus\mathfrak{su}(2)_k^{(1)}\oplus\mathbb{T}^4\ .
\end{equation}

\noindent The so-called ``tensionless string'' is identified with the limit in which the level $k$ is taken to be minimal, i.e. $k=1$. However, the description of the worldsheet theory in the RNS formalism breaks down for the tensionless limit, due to the decomposition \cite{Ferreira:2017}

\begin{equation}
\begin{split}
\mathfrak{sl}(2,\mathbb{R})^{(1)}_k&\cong\mathfrak{sl}(2,\mathbb{R})_{k+2}\oplus\left(3\text{ free Fermions}\right)\\
\mathfrak{su}(2)^{(1)}_k&\cong\mathfrak{su}(2)_{k-2}\oplus\left(3\text{ free Fermions}\right)\ ,
\end{split}
\end{equation}

\noindent Thus, at $k=1$, the $\mathfrak{su}(2)_{k-2}$ factor is non-unitary. Thus, if one wants to study the worldsheet theory in the tensionless limit, one needs to look for an alternative formalism in which the worldsheet theory is well-defined. This is accomplished by the hybrid formalism of Berkovits, Vafa and Witten \cite{Berkovits_1999}, in which one replaces the factor $\rm{AdS}_3\times\rm{S}^3$ with its supergroup analogue $\mathfrak{psu}(1,1|2)_k$, which remains well-defined at level $k=1$.

As was noted in \cite{Eberhardt:2018} and further explored in \cite{Dei:2020}, the hybrid formalism in the tensionless limit has a particularly simple construction in terms of free fields living on the worldsheet. In particular, one constructs the $\mathfrak{u}(1,1|2)_1$ WZW model from four two pairs of free ``symplectic'' Bosons \cite{Goddard:1987} and two pairs of free Fermions, and then quotients out two $\widehat{\mathfrak{u}(1)}$ subalgebras to recover the model on $\mathfrak{psu}(1,1|2)_1$. (The nature of this quotient is a little subtle, and is explained in detail in Section 2 of \cite{Dei:2020}. The exact details, however, will not be relevant here.) As was demonstrated in \cite{Dei:2020}, this free field construction of the $k=1$ theory is an extremely powerful tool for constraining correlation functions of spectrally-flowed highest-weight states on the worldsheet. We will use this construction in this analysis as well to constrain correlation functions at higher genus. In this section, we review the free field construction of the $k=1$ theory, defining the spectrally flowed states and their correlation functions.

\subsection[The free field construction of \texorpdfstring{$\mathfrak{sl}(2,\mathbb{R})_1\oplus\widehat{\mathfrak{u}(1)}$}{sl(2,R)+u(1)}]{\boldmath The free field construction of \texorpdfstring{$\mathfrak{sl}(2,\mathbb{R})_1\oplus\widehat{\mathfrak{u}(1)}$}{sl(2,R)+u(1)}}

As noted above, the basis of our analysis will be the free field construction of the $\mathfrak{u}(1,1|2)_1$ algebra in terms of two pairs of canonically conjugate symplectic Bosons $\xi^{\pm}$ and $\eta^{\pm}$, along with four corresponding Fermions. However, just as in \cite{Dei:2020}, we are primarily concerned with states that are highest-weight with respect to the Fermionic subalgebra. This is essentially equivalent to ignoring the Fermions altogether, and thus we focus on the purely Bosonic part of the free field construction of $\mathfrak{u}(1,1|2)_1$. In terms of the geometry of $\rm{AdS}_3\times\rm{S}^3$, this corresponds to ignoring excitations along the $\rm{S}^3$ factor. In terms of the worldsheet theory, this corresponds to taking the subalgebra $\mathfrak{sl}(2,\mathbb{R})_1\oplus\widehat{\mathfrak{u}(1)}\subset\mathfrak{u}(1,1|2)_1$.\footnote{Strictly speaking, the symplectic Bosons generate a larger chiral algebra which is a conformal extension of $\mathfrak{sl}(2,\mathbb{R})_1\oplus\widehat{\mathfrak{u}(1)}$.} Just as we quotient two $\widehat{\mathfrak{u}(1)}$ algebras from the full $\mathfrak{u}(1,1|2)_1$ model to obtain the $\mathfrak{psu}(1,1|2)_1$ model, we also view the $\widehat{\mathfrak{u}(1)}$ algebra as 'unphysical' and will gauge it to define physical correlation functions.\footnote{The process of 'gauging' the $\widehat{\mathfrak{u}(1)}$ algebra is a little subtle, and is spelled out in a schematic form in Appendix \ref{sec:u-dependence}.} This construction is spelled out in detail in \cite{Dei:2020}, but we review it here for the sake of completeness. Note that, just as in \cite{Dei:2020}, we will focus entirely on the chiral (holomorphic) part of the algebra.

We consider four Bosonic chiral (holomorphic) fields $\xi^{\pm}(z)$ and $\eta^{\pm}(z)$ of conformal weight $h=\frac{1}{2}$ satisfying the OPEs

\begin{equation}
\xi^\alpha(z)\,\eta^\beta(w)\sim\frac{\varepsilon^{\alpha\beta}}{z-w}\ ,
\end{equation}

\noindent while the $\xi\xi$ and $\eta\eta$ OPEs are regular. Here, we have used the convention $\varepsilon^{+-}=-\varepsilon^{-+}=1$. In terms of the mode expansions, this is equivalent to the algebra

\begin{equation}\label{eq:symplectic-Boson-algebra}
[\xi^{\alpha}_r,\eta^{\beta}_{s}]=\varepsilon^{\alpha\beta}\delta_{r+s,0}\ ,
\end{equation}

\noindent where $r,s$ are integer-moded in the R-sector and half-integer moded in the NS-sector. The $(\xi^{\alpha}\eta^{\beta})$ bilinears form $\mathfrak{sl}(2,\mathbb{R})$ currents $J^a(z)$ and a $\mathfrak{u}(1)$ current $U(z)$ given explicitly by

\begin{equation}\label{eq:boson-biliniars}
\begin{gathered}
J^3(z)=-\frac{1}{2}(\xi^+\eta^-)(z)-\frac{1}{2}(\xi^-\eta^+)(z)\ ,\quad J^{\pm}=(\xi^{\pm}\eta^{\pm})(z)\ ,\\
U(z)=\frac{1}{2}(\xi^+\eta^-)(z)-\frac{1}{2}(\xi^-\eta^+)(z)\ .
\end{gathered}
\end{equation}

\noindent A simple calculation shows that the modes of these currents obey the $\mathfrak{sl}(2,\mathbb{R})_1\oplus\widehat{\mathfrak{u}(1)}$ algebra, namely

\begin{equation}
\begin{split}
[J^3_{m},J^3_{n}]=-\frac{1}{2}m\delta_{m+n,0}\ ,&\quad [J^3_{m},J^{\pm}_{n}]=\pm J^{\pm}_{m+n}\ ,\\
[J^+_{m},J^-_{n}]=m\delta_{m+n,0}-2J^3_{m+n}\ ,&\quad [U_{m},U_{n}]=-\frac{1}{2}m\delta_{m+n,0}\ .
\end{split}
\end{equation}

\noindent with all other brackets vanishing. The stress-energy tensor of this theory can be derived from the Sugawara construction, giving

\begin{equation}\label{eq:sugawara-energy-tensor}
T(z)=-\frac{1}{2}\left(J^+J^-+J^-J^+-2J^3J^3\right)(z)-(UU)(z)\ .
\end{equation}

\noindent Alternatively, the stress tensor can be written in terms of free fields

\begin{equation}\label{eq:free-field-energy-tensor}
T(z)=\frac{1}{2}\varepsilon_{\alpha\beta}(\xi^{\alpha}\partial\eta^{\beta}+\eta^{\alpha}\partial\xi^{\beta})(z)\ .
\end{equation}

\noindent The equivalence of \eqref{eq:sugawara-energy-tensor} and \eqref{eq:free-field-energy-tensor} follows from substituting \eqref{eq:boson-biliniars} into \eqref{eq:sugawara-energy-tensor}, and using arguments similar to that given in Section 15.5 of \cite{DiFrancesco} for the free Fermion construction of $\mathfrak{so}(n)_1$.

\subsection{Highest-weight representations}

In the NS-sector of \eqref{eq:symplectic-Boson-algebra}, the symplectic Bosons are half-integer moded and the highest-weight representation is simply given by a single state $\ket{0}$ such that $\xi^{\pm}_{r}\ket{0}=\eta^{\pm}_r\ket{0}=0$ for $r>0$. This defines the trivial representation of $\mathfrak{sl}(2,\mathbb{R})_1\oplus\widehat{\mathfrak{u}(1)}$.

In the R-sector we can construct a highest-weight representation of the symplectic Boson algebra algebra by considering states which are annihilated by $\xi^{\pm}_{m}$ and $\eta^{\pm}_{m}$ for $m>0$, as well as their descendants. The zero modes $J^3_0$ and $U_0$ form a commuting subalgebra, and thus we can label our states by their eigenvalues. That is, we consider states of the form $\ket{m,j}$ such that\footnote{In \cite{Dei:2020}, states were labeled by the values $m_1=(j+m)/2$ and $m_2=(j-m)/2$. This is convenient for explicit computations, in which the modes $\xi_0^{\pm},\eta_0^{\pm}$ act on only one label at a time. Here, however, we choose to label the states by the $J^3_0$ and $U_0$ eigenvalues, since they correspond to the more familiar $\mathfrak{sl}(2,\mathbb{R})$ labels.}

\begin{equation}
\xi_{m}^{\pm}\ket{m,j}=\eta^{\pm}_{m}\ket{m,j}=0\ ,\quad m>0
\end{equation}

\noindent and which also satisfy

\begin{equation}
J^3_0\ket{m,j}=m\ket{m,j}\ ,\quad U_0\ket{m,j}=(j-\tfrac{1}{2})\ket{m,j}\ .
\end{equation}

\noindent Such a representation was constructed in \cite{Eberhardt:2018} and later in \cite{Dei:2020}, and the zero modes of $\xi^{\pm}$ and $\eta^{\pm}$ act on these states as

\begin{equation}
\begin{split}
\xi^{+}_0\ket{m,j}=\ket{m+\tfrac{1}{2},j-\tfrac{1}{2}}\ ,&\quad \eta^+_0\ket{m,j}=(m+j)\ket{m+\tfrac{1}{2},j+\tfrac{1}{2}}\ ,\\
\xi^{-}_0\ket{m,j}=-\ket{m-\tfrac{1}{2},j-\tfrac{1}{2}}\ ,&\quad \eta^-_0\ket{m,j}=-(m-j)\ket{m-\tfrac{1}{2},j+\tfrac{1}{2}}\ .
\end{split}
\end{equation}

\noindent The convention for this representation is chosen so that the $\mathfrak{sl}(2,\mathbb{R})$ currents act as

\begin{equation}
J^{\pm}_0\ket{m,j}=(m\pm j)\ket{m\pm 1,j}\ .
\end{equation}

\noindent Finally, the quadratic Casimir of the $\mathfrak{sl}(2,\mathbb{R})$ zero-mode algebra is simply given by

\begin{equation}
\mathcal{C}^{\mathfrak{sl}(2,\mathbb{R})}\ket{m,j}=j(1-j)\ket{m,j}\ ,
\end{equation}

\noindent which justifies us in calling $j$ the ``spin'' of the associated $\mathfrak{sl}(2,\mathbb{R})$ representation. 

Since $\xi^{\pm}_0$ lowers $j$ by $\frac{1}{2}$ and $\eta^{\pm}_0$ raises $j$ by $\frac{1}{2}$, a generic irreducible representation of the symplectic Boson algebra \eqref{eq:symplectic-Boson-algebra} contains many irreducible representations of $\mathfrak{sl}(2,\mathbb{R})_1$. However, as noted above, in order to relate the free field states to physical states on $\rm{AdS}_3$, we must gauge the $\widehat{\mathfrak{u}(1)}$ subalgebra generated by $U(z)$. In terms of highest-weight states, this means keeping only those states with $j=\frac{1}{2}$ (which agrees with the representation theory of $\mathfrak{psu}(1,1|2)_1$ \cite{Eberhardt:2018}). However, when we define correlation functions, we will find that the picture changing operators we insert do not commute with $U_0$, and thus change the value of $j$ of the corresponding state on which they act. Because of this, quotienting out the $\widehat{\mathfrak{u}(1)}$ subalgebra is more subtle than just picking states with $j=\frac{1}{2}$. Thus, we will keep the label $j$ arbitrary, with the understanding that we can set $j$ to $\frac{1}{2}$ at the end of the analysis.

\subsection{Spectral flow}

The symplectic Boson algebra contains a family of outer automorphisms generated by two ``spectral flow'' operations which we will call $\sigma^{(\pm)}$. Specifically, these automorphisms act on the modes of \eqref{eq:symplectic-Boson-algebra} via

\begin{equation}
\begin{split}
\sigma^{(+)}(\xi^-_{r})=\xi^-_{r+\frac{1}{2}}\ ,&\quad\quad\quad\sigma^{(-)}(\xi^+_r)=\xi^+_{r-\frac{1}{2}}\ ,\\
\sigma^{(+)}(\eta^+_{r})=\eta^+_{r-\frac{1}{2}}\ ,&\quad\quad\quad\sigma^{(-)}(\eta^-_r)=\eta^-_{r+\frac{1}{2}}\ .
\end{split}
\end{equation}

A proper analysis of the worldsheet theory on $\rm{AdS}_3$ requires not just the highest-weight representations mentioned above, but also their images under spectral flow. Instead of parameterizing the spectral flow automorphisms in terms of $\sigma^{(\pm)}$, one can consider the mixed spectral flow operators

\begin{equation}
\sigma\equiv\sigma^{(+)}\circ\sigma^{(-)}\ ,\quad\widehat{\sigma}\equiv\sigma^{(+)}\circ(\sigma^{(-)})^{-1}\ .
\end{equation}

\noindent The spectral flow automorphisms $\sigma$ and $\widehat{\sigma}$ act independently on the $\mathfrak{sl}(2,\mathbb{R})_1$ and $\widehat{\mathfrak{u}(1)}$ currents, respectively. In particular, we have

\begin{equation}
\sigma^w(J_n^\pm)=J^{\pm}_{n\mp w}\ ,\quad \sigma^w(J^3_n)=J^3_n+\frac{w}{2}\delta_{n,0}\ ,\quad \widehat{\sigma}^w(U_n)=U_n+\frac{1}{2}\delta_{n,0}\ ,
\end{equation}

\noindent and the action on all other fields is trivial.

Given a state $\ket{\psi}$ in some representation of the free field algebra, we can also define the spectrally flowed state $\ket{\psi}^{(1,0)}$ implicitly via the relation\footnote{Note that we could, instead, first define the spectral flow $\Sigma:\mathcal{H}\to\mathcal{H}$ acting on states, and then \eqref{eq:spectrally-flowed-states} is simply the statement that $\Sigma$ defines an autmorphism on the space of linear operators defined by conjugation with $\Sigma$, i.e. $\sigma(A)=\Sigma^{-1}A\Sigma$.}

\begin{equation}\label{eq:spectrally-flowed-states}
A\ket{\psi}^{(1,0)}=(\sigma(A)\ket{\psi})^{(1,0)}\ .
\end{equation}

\noindent More generally, we can define a state $\ket{\psi}^{(p,q)}$ which is spectrally flowed $p$ times with $\sigma$ and $q$ times with $\widehat{\sigma}$ which satisfies\footnote{It should be noted that this is a slightly different notation than what was used in \cite{Dei:2019}.}

\begin{equation}
\begin{split}
A\ket{\psi}^{(p,q)}=\left(\sigma^p\circ\widehat{\sigma}^q(A)\ket{\psi}\right)^{(p,q)}\ .
\end{split}
\end{equation}

\noindent Central to our analysis are the spectrally flowed images of ground states $\ket{m,j}$. The action of the Boson modes on these states is given by

\begin{equation}\label{eq:spectrally-flowed-states-action}
\begin{split}
\xi_r^{+}\ket{m,j}^{(p,q)}=\left(\xi^{+}_{r-\frac{p-q}{2}}\ket{m,j}\right)^{(p,q)}=0\ ,&\quad r>\frac{p-q}{2}\ ,\\
\xi_r^{-}\ket{m,j}^{(p,q)}=\left(\xi^{-}_{r+\frac{p+q}{2}}\ket{m,j}\right)^{(p,q)}=0\ ,&\quad r>-\frac{p+q}{2}\ ,\\
\eta_r^{+}\ket{m,j}^{(p,q)}=\left(\eta^{+}_{r-\frac{p+q}{2}}\ket{m,j}\right)^{(p,q)}=0\ ,&\quad r>\frac{p+q}{2}\ ,\\
\eta_r^{-}\ket{m,j}^{(p,q)}=\left(\eta^{-}_{r+\frac{p-q}{2}}\ket{m,j}\right)^{(p,q)}=0\ ,&\quad r>-\frac{p-q}{2}\ .
\end{split}
\end{equation}

\subsection{Vertex operators}

We now define the vertex operators which we will use to calculate correlation functions. Given the state $\ket{m,j}^{(p,q)}$, i.e. the spectrally flowed image of a highest-weight state of the symplectic Boson algebra, we can define the vertex operator $V_{m,j}^{(p,q)}(z)$ to be the local operator associated to that state. Since $\widehat{\sigma}$ spectrally flows the unphysical $\widehat{\mathfrak{u}(1)}$ subalgebra generated by $U(z)$, the states which correspond to string states on $\rm{AdS}_3$ are those which are only flowed by $\sigma$. These will be the states whose correlation functions we are interested in, and will be given the shorthand

\begin{equation}
V^{w}_{m,j}(z):=V^{(w,0)}_{m,j}(z)\ .
\end{equation}

The vertex operators $V^{w}_{m,j}(z)$ give us a local operator associated to the insertion of a string state at the point $z$ on the worldsheet $\Sigma$. It will also be useful to give the vertex operator a label $x$ associated to the point on the dual CFT spacetime. Since the holographic dictionary tells us that $J^+_0$ acts as the translation operator on the dual CFT sphere \cite{Maldacena-Ooguri-1}, it makes sense to introduce an extra coordinate $x$ on our vertex operators, labeling the dual CFT coordinate, by

\begin{equation}
V^{w}_{m,j}(x;z)=e^{xJ^+_0}V^{w}_{m,j}(z)e^{-xJ_0^+}\ .
\end{equation}

Finally, we need the OPEs of $\xi^{\pm}(y)$ with $V^{w}_{m,j}(x;z)$.\footnote{One could write down the OPEs with $\eta^{\pm}(z)$ as well, but we will not need them here.} By \eqref{eq:spectrally-flowed-states-action}, we have

\begin{equation}\label{eq:OPEs}
\begin{split}
\xi^+(y)V^{w}_{m,j}(z)&=(y-z)^{-\frac{w+1}{2}}V^{w}_{m+\frac{1}{2},j+\frac{1}{2}}(z)+\mathcal{O}\left((y-z)^{-\frac{w-1}{2}}\right)\ ,\\
\xi^-(y)V^{w}_{m,j}(z)&=(y-z)^{\frac{w-1}{2}}V^{w}_{m-\frac{1}{2},j+\frac{1}{2}}(z)+\mathcal{O}\left((y-z)^{\frac{w+1}{2}}\right)\ .
\end{split}
\end{equation}

\noindent From here, one can find the OPE of $\xi^{\pm}(y)$ with $V_{m,j}^{w}(x;z)$ by noting that

\begin{equation}
\xi^{\pm}(y)V_{m,j}^{w}(x;z)=e^{xJ^+_0}\xi^{\pm(x)}(y)V_{m,j}^{w}(z)e^{-xJ^+_0\ ,}
\end{equation}

\noindent where

\begin{equation}\label{eq:xi-x}
\begin{split}
\xi^{+(x)}(y)=e^{-xJ^+_0}\xi^+(y)e^{xJ^+_0}&=\xi^+(y)\ ,\\
\xi^{-(x)}(y)=e^{-xJ^+_0}\xi^-(y)e^{xJ^+_0}&=\xi^-(y)-x\,\xi^+(z)\ .
\end{split}
\end{equation}

\noindent A consequence is that the combination $\xi^-(y)+x\,\xi^+(y)$ has a regular OPE with $V_{m,j}^{w}(x;z)$, since by \eqref{eq:OPEs} and \eqref{eq:xi-x}, we have

\begin{equation}\label{eq:regular-OPE}
\begin{split}
\left(\xi^-(y)+x\,\xi^+(y)\right)V^{w}_{m,j}(x;z)&=e^{xJ_0^+}\xi^-(y)V_{m,j}^{w}(x;z)e^{xJ_0^+}\\
&=(y-z)^{\frac{w-1}{2}}V^{w}_{m-\frac{1}{2},j+\frac{1}{2}}(x;z)+\mathcal{O}\left((y-z)^{\frac{w+1}{2}}\right)\ .
\end{split}
\end{equation}

\noindent The regularity of this OPE provides enormous constraints on the correlation functions of the spectrally flowed ground states, and, as we will see in Section \ref{sec:Incidence-relation}, essentially fixes the correlation functions to be those of the dual CFT.

\subsection{Picture changing and correlation functions}

Now that we have defined the fields, states and vertex operators of the worldsheet theory, we move to defining the correlation functions of spectrally flowed ground states. As we have previously noted, we are only interested in studying the chiral (or anti-chiral) correlators. A full analysis of the worldsheet theory would involve mixed correlators with both holomorphic and anti-holomorphic dependence. In order to fully determine such correlation functions, one would need to solve the conformal bootstrap equations for tensionless string, which, to the knowledge of this author, has not yet been done. 

In order to study the correlation functions of the vertex operators $V_{m_i,j_i}^{w_i}(x_i;z_i)$, the naive approach would be to simply study the correlators

\begin{equation}\label{eq:naive-correlator}
\Bigl\langle\prod_{i=1}^{n}V_{m_i,j_i}^{w_i}(x_i;z_i)\Bigr\rangle\ .
\end{equation}

\noindent However, as noted in \cite{Dei:2020}, the naive answer needs to be modified, since in the hybrid formalism of \cite{Berkovits_1999}, correlation functions are defined with the insertion of extra operators. In the present scenario, this modification amounts to inserting operators analogous to picture changing operators familiar in the RNS formalism. The form of the picture changing operator in general is rather complicated (involving free Fermions and ghosts), but because we consider states which are highest weight with respect to the Fermions of the $\mathfrak{psu}(1,1|2)_1$ model, the picture changing operators are essentially given by $[\xi^+_0\xi^-_0,\cdot]$,\footnote{Note that one could have instead picked the picture changing operator $[\eta^+_0\eta^-_0,\cdot]$. However, this would make no difference in the resulting correlation functions due to the $\mathbb{Z}_2$ automorphism $\xi^{\pm}\leftrightarrow\eta^{\pm}$ of the symplectic Boson algebra.} up to normalization (see Section 3.1 of \cite{Dei:2020}). Since a correlator on a genus $g$ worldsheet requires $n+2g-2$ picture changing operators, we must replace $n+2g-2$  vertex operators\footnote{The number $n+2g-2$ comes from the definition of physical correlators in the hybrid formalism \cite{Berkovits_1999}, which in turn comes from the construction of correlators in topological string theory \cite{Berkovits:1994vy}. However, this is also the number of picture changing operators required in the RNS formalism \cite{Friedan:1985ge}.} with the modified fields\footnote{If $g\geq 2$, there are more picture changing operators than vertex operators. In this case, one can simply act on the same vertex operator more than once. The resulting correlator should be independent of this choice after gauging the unphysical $\widehat{\mathfrak{u}(1)}$ subalgebra of the free field construction.}

\begin{equation}
[\xi^+_0\xi^-_0,V_{m_i,j_i}^{w_i}(x;z)]=V_{m_i,j_i-1}^{w_i}(x;z)\ .
\end{equation}

The next naive guess would be to consider correlation functions of the form \eqref{eq:naive-correlator} with picture changing operators inserted. However, this also runs into a problem. Since we want to define correlation functions for strings in $\rm{AdS}_3$, we must gauge the $\widehat{\mathfrak{u}(1)}$ algebra generated by $U(z)$. At the level of the vertex operators, this corresponds to setting $j_i=\frac{1}{2}$ for each state \textit{before picture changing}. After the application of picture changing operators, $n+2g-2$ of the spins $j_i$ are shifted by one unit. Thus, if we let $\tilde{j}_i$ be the spins after picture changing, we have

\begin{equation}\label{eq:picture-changing-spins}
\sum_{i=1}^{n}\Bigl(\tilde{j}_i-\frac{1}{2}\Bigr)=-(n+2g-2)\ .
\end{equation}

\noindent However, a correlator with such a set of spins vanishes, since the global Ward identities for the current $U(z)$ imply that

\begin{equation}
0=-\oint\frac{\mathrm{d}z}{2\pi i}\Bigl\langle U(z)\prod_{i=1}^{n}V_{m_i,\tilde{j}_i}^{w_i}(x_i;z_i)\Bigr\rangle=\sum_{\ell=1}^{n}\Bigl(\tilde{j}_{\ell}-\frac{1}{2}\Bigr)\Bigl\langle\prod_{i=1}^{n}V_{m_i,\tilde{j}_i}^{w_i}(x_i;z_i)\Bigr\rangle\ ,
\end{equation}

\noindent where the contour is a small circle which does not enclose any point $z_i$. Thus, the correlation function obtained by picture changing also doesn't define a meaningful correlator (unless $n+2g-2=0$).

The resolution proposed in \cite{Dei:2020} is to insert $n+2g-2$ fields into the correlator, each of which maps to the ground state after gauging the $\widehat{\mathfrak{u}(1)}$ subalgebra, and each of which has $U_0$ eigenvalue $1$. Such a state is given by

\begin{equation}
\ket{W}:=\ket{\Omega}^{(0,2)}\ ,
\end{equation}

\noindent where $\ket{\Omega}$ is the (NS sector) ground state of our theory.\footnote{Equivalently, we could have added a single copy of the field $\ket{\Omega}^{(0,2n+4g-4)}$.} If we let $W(u)$ be the vertex operator associated to the field $\ket{W}$, we then define the correlation function of the fields $V_{m_i,j_i}^{w_i}(x_i;z_i)$ to be

\begin{equation}\label{eq:generic-correlator}
\Bigl\langle\prod_{\alpha=1}^{n+2g-2}W(u_{\alpha})\prod_{i=1}^{n}V_{m_i,\tilde{j}_i}^{w_i}(x_i;z_i)\Bigr\rangle\ .
\end{equation}

\noindent This correlation function will be our main object of study throughout the rest of the paper. Since we leave generic the spins $j_i$, we will omit the tildes, with the understanding that we can set $j_i$ to specific values satisfying \eqref{eq:picture-changing-spins} later. It should be noted that, while \eqref{eq:generic-correlator} depends on the locations $u_{\alpha}$ of the auxiliary fields $W(u_{\alpha})$, this is a consequence of the fact that we are working in the free field realization of $\mathfrak{sl}(2,\mathbb{R})_1\oplus\widehat{\mathfrak{u}(1)}$. Once we gauge the $\widehat{\mathfrak{u}(1)}$ subalgebra, the resulting correlation functions will no longer depend on these unphysical insertion points. The dependence of these correlators on $u_{\alpha}$, and the relationship to the physical correlators on $\rm{AdS}_3$, are briefly discussed in Appendix \ref{sec:u-dependence}.

Just as in \cite{Dei:2020}, the OPE of $\xi^{\pm}(z)$ with $W(u)$ will be important in what follows. Since $\ket{W}=\ket{\Omega}^{(0,2)}$ and $\ket{\Omega}$ is annihilated by all positive modes of $\xi^{\pm}$, and since $\xi^{\pm}_{r}\ket{W}=\left(\xi^{\pm}_{r+1}\ket{\Omega}\right)^{(0,2)}$, we see that $\ket{W}$ is annihilated by the modes $\xi^{\pm}_{r}$ with $r>-\frac{3}{2}$. From this, we conclude the OPE

\begin{equation}
\xi^{\pm}(z)W(u)\sim\mathcal{O}(z-u)\ .
\end{equation}

\noindent That is, the OPE of $\xi^{\pm}$ with $W$ has no singular nor constant terms.

\section{The worldsheet as a covering space}\label{sec:Incidence-relation}

We now turn our attention toward the main result, which lies in a remarkable identity relating correlation functions with $\xi^{\pm}(z)$ insertions to the covering map $\Gamma:\Sigma\to\rm{S}^2$ which defines the correlators of twist fields in the dual CFT. This identity immediately implies the localization of the correlation functions \eqref{eq:generic-correlator} to points in the moduli space where such a covering exists, which is the key property that makes the equivalence to correlation functions in the symmetric orbifold manifest. In particular, it concretely realizes the string worldsheet as the covering space used in the calculation of symmetric orbifold correlators. What's more, despite this identity's power and utility, its proof is based on a simple counting of poles and zeroes of certain correlation functions. An analogous identity for genus zero correlation functions was the central point of \cite{Dei:2020}, which we generalize to higher genus worldsheets. We use the definitions and results presented in Appendix \ref{sec:Riemann-Surfaces} extensively.

\subsection{Covering maps}

Before we can analize correlation functions, let us first give a brief review of branched covering maps, using the conventions of \cite{Eberhardt:2019,Eberhardt:2020,Dei:2020}. Consider some holomorphic function $\Gamma:\Sigma\to\rm{S}^2$, where $\rm{S}^2$ is endowed with the complex structure of the Riemann sphere. Given $n$ marked points $z_i$ on $\Sigma$, $n$ marked point $x_i$ on $\rm{S}^2$, and $n$ positive integers $w_i$, we say that $\Gamma$ is a \textit{branched covering map} mapping $z_i$ to $x_i$ with branching index $w_i$ if, near $z_i$, $\Gamma$ has the local behavior

\begin{equation}
\Gamma(z)\sim x_i+a_i^{\Gamma}(z-z_i)^{w_i}+\mathcal{O}\left((z-z_i)^{w_i+1}\right)\ ,
\end{equation}

\noindent where $a_i^{\Gamma}$ is some constant in $z$. Furthermore, we require that $\Gamma$ has no other critical points. 

A central result in the theory of covering maps is the Riemann-Hurwitz formula, which counts how many times the covering space covers the sphere. Specifically, the number $N$ of preimages of any generic point is

\begin{equation}\label{eq:Riemann-Hurwitz}
N=1-g+\sum_{i=1}^{n}\frac{w_i-1}{2}\ .
\end{equation}

\noindent To derive it, note that the derivative $\partial\Gamma$ defines a meromorphic one-form on $\Sigma$ (see \ref{subsec:forms} and \ref{subsec:divisors}). Thus we have $Z(\partial\Gamma)-P(\partial\Gamma)=2g-2$, where $Z(\cdot)$ and $P(\cdot)$ count the number of zeroes and poles of a function (with multiplicity). Since $\partial\Gamma$ has zeroes at $z_i$ of order $w_i-1$, we have

\begin{equation}
P(\partial\Gamma)=2-2g+\sum_{i=1}^{n}(w_i-1)\ .
\end{equation}

\noindent Since a simple pole of $\Gamma$ is a double pole of $\partial\Gamma$, we know that $P(\partial\Gamma)=2P(\Gamma)$, and thus the number of poles of $\Gamma$ (the number of preimages of $\infty$) is given by \eqref{eq:Riemann-Hurwitz}. Since the number of preimages (including multiplicity) of any point on the sphere is the same, the Riemann-Hurwitz formula follows.

An important property of covering maps is that, given a set of data $\{x_i,w_i\}$ of points on $\rm{S}^2$ and branching indices, a covering map only exists for specific values of the insertion points $z_i$ and worldsheet moduli. Indeed, as we will see in Section \ref{sec:Ward-to-covering}, the covering map $\Gamma$ can be constructed by a set of $2N+n+3g-2$ algebraic constraints in $2N+1$ variables. In order for a solution to exist, $n+3g-3$ sets of relations between the worldsheet moduli must be satisfied. Since the dimension of the moduli space of a Riemann surface of genus $g$ with $n$ marked points is $\dim{\mathcal{M}_{g,n}}=n+3g-3$, one would expect that the moduli which allow for a covering map form a discrete set on the moduli space. This, as it turns out, is the case, and this is essentially the mechanism by which the worldsheet correlation functions localize to discrete points in the string moduli space (see the discussion in Section 5.1.2 of \cite{Eberhardt:2020}).

\subsection{The incidence relation}

Now that we have defined the covering map and a few of its properties, we can turn to the central claim of this paper. Let $\{z_i\in\Sigma\}$ and $\{x_i\in\rm{S}^2\}$ be points on the worldsheet and the Riemann sphere, and let $\{w_i\}$ be a set of branching indices. If there exists a covering map $\Gamma:\Sigma\to\rm{S}^2$ mapping $z_i$ to $x_i$ with branching index $w_i$, then the identity

\begin{equation}\label{eq:incidence-relation}
\Bigl\langle\left(\xi^-(z)+\Gamma(z)\xi^+(z)\right)\prod_{\alpha=1}^{n+2g-2}W(u_{\alpha})\prod_{i=1}^{n}V_{m_i,j_i}^{w_i}(x_i;z_i)\Bigr\rangle=0
\end{equation}

\noindent holds for all values of $z$, $m_i$ and $j_i$. Moreover, if such a covering does not exist, then the correlation function \eqref{eq:generic-correlator} vanishes identically.

As pointed out in \cite{Dei:2020}, \eqref{eq:incidence-relation} is reminiscent of the identity $\mu_{\dot{a}}+x_{a\dot{a}}\lambda^a=0$ in twistor theory, called the ``incidence relation,'' relating the bispinor $(\mu_{\dot{a}},\lambda_a)$ to the spacetime coordinates $x_{a\dot{a}}=\sigma^{\mu}_{a\dot{a}}x_{\mu}$. In \eqref{eq:incidence-relation}, the role of the bispinors are played by $\xi^-$ and $\xi^+$, and the role of the Minkowski coordinate is played by the dual CFT coordinate $x=\Gamma(z)$. This hints at a potential twistor description of tensionless strings on $\rm{AdS}_3$, similar to the twistor open string theory proposed for $\mathcal{N}=4$ super Yang-Mills \cite{Berkovits_2004}.

\subsection{Localization}

Before we move to the proof of \eqref{eq:incidence-relation}, let us first discuss its consequences. The first and most striking is that the correlation functions \eqref{eq:generic-correlator} localize on the worldsheet moduli space $\mathcal{M}_{g,n}$ to those points which admit a branched covering to $x_i$ with branching indices $w_i$. While there are many distributions which localize to discrete spaces in the moduli space, it is natural to assume that these functions take the form of delta functions. Thus, the correlators apparently take the form

\begin{equation}\label{eq:localization}
\Bigl\langle\prod_{\alpha=1}^{n+2g-2}W(u_{\alpha})\prod_{i=1}^{n}V_{m_i,j_i}^{w_i}(x_i;z_i)\Bigr\rangle=\sum_{\Gamma}C_{\Gamma}\prod_{\ell=1}^{n+3g-3}\delta(f_{\ell}^{\Gamma})\ ,
\end{equation}

\noindent for some function $C_{\Gamma}$ of the parameters. The sum here is over the (finite number of) covering maps permitted for a choice of $x_i,w_i$, and the $f_{\ell}^{\Gamma}$ are a set of constraints which, when set to zero, uniquely determine the point on $\mathcal{M}_{g,n}$ on which $\Gamma$ exists. This localization tells us that, at the level of string perturbation theory, only worldsheets which are covering spaces of the sphere with marked points $x_i$ and branching indices $w_i$ contribute to any physical scattering process. Since $\Gamma$ here is exactly the covering map used to calculate the correlators in the symmetric orbifold theory \cite{Lunin_2001}, this very concretely realizes the conjecture that, in the string theory dual to the symmetric orbifold, the worldsheet has the interpretation of that covering space \cite{Pakman_2009}.

While the localization of the worldsheet correlation functions to specific points in the moduli space is unusual in most string theory models, it is commonplace in models of topological strings \cite{Verlinde:1990ku,Belavin:2006ex}. Furthermore, it is expected that the worldsheet duals to free gauge theories also have correlation functions which localize to some lower-dimensional hypersurface in the moduli space \cite{Aharony:2006th,Aharony:2007fs,Razamat:2008zr}. Thus, since the tensionless limit of $\rm{AdS}_3$ string theory is expected to be ``topological'' in some sense, and since the symmetric orbifold shares many properties with free gauge theories \cite{Pakman_2009}, the localization of these correlation functions is less surprising than one would initially expect.

\subsection{Non-renormalization}

The localization property of correlation functions of spectrally flowed highest-weight states has another consequence, as was noted in \cite{Eberhardt:2019,Eberhardt:2020}. In string perturbation theory, we schematically define the perturbative amplitude of the scattering of string states with asymptotic winding $w_i$ emanating from the point $x_i$ on the $\rm{AdS}_3$ boundary to be

\begin{equation}\label{eq:perturbation-theory}
\mathcal{A}(x_i,w_i)=\sum_{g=0}^{\infty}g_s^{2-2g}\int_{\mathcal{M}_{g,n}}\mathrm{d}\mu\,\Bigl\langle\prod_{\alpha=1}^{n+2g-2}W(u_{\alpha})\prod_{i=1}^{n}V_{m_i,j_i}^{w_i}(x_i;z_i)\Bigr\rangle\ ,
\end{equation}

\noindent where the correlation function is computed on a Riemann surface of genus $g$. However, as we have shown, the correlation functions are only nonzero if a covering $\Gamma:\Sigma\to\rm{S}^2$ exists with branching $w_i$. Furthermore, for a fixed set of branching indices $w_i$, there is a maximal genus for which a covering map exists.\footnote{This is a simple consequence of the Riemann-Hurwitz formula. Indeed, we have that $N\geq w_i$ for all $i$. Otherwise, the point $x_i$ would have more than $N$ preimages, counting multiplicity, contradicting the definition of $N$. We can use the Riemann-Hurwitz formula \eqref{eq:Riemann-Hurwitz} to rewrite this inequality as

\begin{equation*}
g\leq \sum_{j\neq i}^{n}\frac{w_j-1}{2}-\frac{w_i-1}{2}\ ,
\end{equation*}

\noindent and thus the genus $g$ cannot be arbitrarily large for a given set of branching indices $w_i$.} It follows that, for any amplitude $\mathcal{A}(x_i,w_i)$, the perturbation series \eqref{eq:perturbation-theory} truncates to a finite sum. This non-renormalization theorem was proposed in \cite{Eberhardt:2019} and further expanded upon in \cite{Eberhardt:2020}. However, in those works, the non-renormalization was shown only for a particular set of solutions to the worldsheet Ward identities. Here, we have explicitly demonstrated that the perturbation series truncates, at least for correlation functions of spectrally flowed highest-weight states.

\subsection{Proof of the incidence relation}

We now turn our head toward proving \eqref{eq:incidence-relation} and the corresponding localization principle. To this end, we define $1/2$-forms\footnote{Implicitly, defining the spinors $\omega^{\pm}$ requires a choice of spin structure on $\Sigma$. However, our analysis is independent of the exact spin structure chosen, and so we keep it arbitrary. See Appendix \ref{sec:u-dependence} for a short discussion on this matter.} on $\Sigma$ (see Section \ref{subsec:spinors}) by

\begin{equation}
\omega^{\pm}(z)=\Bigl\langle\xi^{\pm}(z)\prod_{\alpha=1}^{n+2g-2}W(u_{\alpha})\prod_{i=1}^{n}V_{m_i,j_i}^{w_i}(x_i;z_i)\Bigr\rangle\ .
\end{equation}

\noindent Now, note that $\omega^+$ and $\omega^-$ have branch points at the same locations -- namely, the locations $z=z_i$ with $w_i$ even. Thus, assuming $\omega^+$ doesn't identically vanish, the ratio $-\omega^-(z)/\omega^+(z)$ is a globally defined meromorphic function with no branch cuts. That is, we can implicitly define a merormophic function $\gamma:\Sigma\to\rm{S}^2$ by the relation

\begin{equation}\label{eq:gamma-implicit-def}
\omega^-(z)+\gamma(z)\,\omega^+(z)=0\ .
\end{equation}

\noindent Our goal is now to show that $\gamma(z)$ is precisely the covering map, when such a map exists. First, let us note that, by \eqref{eq:regular-OPE}, we have

\begin{equation}\label{eq:omega-regularity}
\omega^-(z)+x_i\,\omega^+(z)\sim\mathcal{O}\left((z-z_i)^{\frac{w_i-1}{2}}\right)\ .
\end{equation}

\noindent We can subtract \eqref{eq:omega-regularity} from \eqref{eq:gamma-implicit-def} to get

\begin{equation}
(\gamma(z)-x_i)\,\omega^+(z)\sim\mathcal{O}\left((z-z_i)^{\frac{w_i-1}{2}}\right)\ .
\end{equation}

\noindent Since $\omega^+(z)$ has a pole of order $\frac{w_i+1}{2}$ at $z=z_i$ (by the OPEs \eqref{eq:OPEs}), we conclude that $\gamma(z_i)=x_i$ and that $\gamma(z)$ has a critical point of order $w_i$ at $z_i$, i.e.

\begin{equation}
\gamma(z)-x_i\sim\mathcal{O}\left((z-z_i)^{w_i}\right)\ ,\quad z\to z_i\ .
\end{equation}

\noindent Therefore, $\gamma(z)$ is \textit{some} ramified covering of $\Sigma$ by the sphere with critical points of order $w_i$ at $z=z_i$. We will be able to show that $\gamma(z)$ is \textit{the} covering map in question if we can show that these are all of its critical points. To do this, we show that $\gamma$ has order $N$, given by \eqref{eq:Riemann-Hurwitz}, since if $\gamma$ had more critical points, its order would be greater than $N$. 

Let $P(\gamma)$ be the number of poles of $\gamma$, counting multiplicity. Then $P(\gamma)$ is also the order of $\gamma$, which by \eqref{eq:Riemann-Hurwitz}, must be at least $N$. That is,

\begin{equation}\label{eq:inequality-geq}
P(\gamma)\geq N= 1-g+\sum_{i=1}^{n}\frac{w_i-1}{2}\ .
\end{equation} 

\noindent To show that $P(\gamma)$ has no other critical points, it suffices to show that this bound is saturated. We do this by explicitly counting the poles of $\gamma$ in terms of the zeroes of $\omega^+$ and the poles of $\omega^-$.

Since $\gamma(z)=-\omega^-(z)/\omega^+(z)$, any pole of $\gamma$ comes from either a pole in $\omega^-$ or a zero in $\omega^+$. Since $\omega^-$ and $\omega^+$ share all of the same poles, with the same multiplicity, we conclude that poles of $\gamma$ can only come from zeroes in $\omega^+$. Since $\omega^+$ is a meromorphic $1/2$-form, the number of zeroes of $\omega^+$ is related to the number of poles by $Z(\omega^+)-P(\omega^+)=g-1$, and thus

\begin{equation}
Z(\omega^+)=P(\omega^+)+g-1=\sum_{i=1}^{n}\frac{w_i+1}{2}+g-1\ .
\end{equation}

\noindent Naively, this yields the number of poles of $\gamma$. However, since $\xi^{\pm}(z)W(u_{\alpha})\sim\mathcal{O}(z-u_{\alpha})$, $\omega^-$ and $\omega^+$ also share $n+2g-2$ simple zeroes at $z=u_{\alpha}$, and thus $\gamma(z)$ has removable singularities at $z=u_{\alpha}$, and these points do not contribute to $P(\gamma)$. Therefore, the number of poles of $\gamma$ is constrained by

\begin{equation}\label{eq:inequality-leq}
P(\gamma)\leq Z(\omega^+)-(n+2g-2)=1-g+\sum_{i=1}^{n}\frac{w_i-1}{2}=N\ .
\end{equation}

\noindent Equations \eqref{eq:inequality-geq} and \eqref{eq:inequality-leq} are only simultaneously satisfied if $P(\gamma)=N$, and thus $\gamma$ has no other critical points, and must be the covering map.

Now, if a covering map does not exist, we must have $\omega^+(z)=0$ identically, since otherwise the ratio $-\omega^-(z)/\omega^+(z)$ would define such a covering map by the argument above. This implies that the correlation functions \eqref{eq:generic-correlator} vanish identically. To see this, note that, by the OPE of $\xi^+$ with $V^{w}_{m,j}$ (see \eqref{eq:OPEs}), we have

\begin{equation}
\omega^+(z)\sim\frac{1}{(z-z_i)^{\frac{w_i+1}{2}}}\Bigl\langle\prod_{\alpha=1}^{n+2g-2}W(u_{\alpha})\,V_{m_i+\frac{1}{2},j_i+\frac{1}{2}}^{w_i}(x_i;z_i)\prod_{j\neq i}^{n}V_{m_j,j_j}^{w_j}(x_j;z_j)\Bigr\rangle+\cdots\ .
\end{equation}

\noindent Thus, if $\omega^+(z)$ identically vanishes, we must have

\begin{equation}
\Bigl\langle\prod_{\alpha=1}^{n+2g-2}W(u_{\alpha})\,V_{m_i+\frac{1}{2},j_i+\frac{1}{2}}^{w_i}(x_i;z_i)\prod_{j\neq i}^{n}V_{m_j,j_j}^{w_j}(x_j;z_j)\Bigr\rangle=0\ .
\end{equation}

\noindent Simply shifting the indices $m_i,j_i$ (which we kept arbitrary), we find that the correlation function vanishes unless a covering map exists. This proves the localization behavior of the correlators.

\subsection{Constraining the correlators}

Apart from demonstrating the localization properties of correlators and the non-renormalization of the correlation functions, the incidence relation \eqref{eq:incidence-relation} can be used to almost fully constrain the form of the correlation functions \eqref{eq:generic-correlator}. In particular, we can fully constrain the dependence of the correlators on the $\mathfrak{sl}(2,\mathbb{R})$ index $m_i$. This is essentially exactly the same calculation done in \cite{Dei:2020}, but we replicate it here for completeness.

We begin by recalling the behavior of the covering map $\Gamma(z)$ near $z=z_i$, defined above. Near $z=z_i$, we have

\begin{equation}
\Gamma(z)\sim x_i+a_i^{\Gamma}(z-z_i)^{w_i}+\mathcal{O}\left((z-z_i)^{w_i+1}\right)\ ,\qquad z\to z_i\ ,
\end{equation}

\noindent where $a_i^{\Gamma}$ are some constant in $z$, which will generically depend on the data $x_i,w_i$. The trick now is to assume the configuration $\{z_i,x_i,w_i\}$ is such that a covering map exists, and to expand the identity \eqref{eq:incidence-relation} around the point $z=z_i$ by using the local behavior of $\Gamma$, as well as the OPEs \eqref{eq:OPEs}. Taking this all into account, we have

\begin{equation}
\begin{split}
&\Bigl\langle \left(\xi^-(z)+\Gamma(z)\, \xi^+(z)\right)\prod_{\alpha=1}^{n+2g-2}W(u_{\alpha})\prod_{j=1}^{n}V_{m_j,j_j}^{w_j}(x_j;z_j)\Bigr\rangle\\
&\hspace{2cm}\sim -(z-z_i)^{\frac{w_i-1}{2}}\Bigl\langle \prod_{\alpha=1}^{n+2g-2}W(u_{\alpha})\, V_{m_i-\frac{1}{2},j_i-\frac{1}{2}}^{w_i}(x_i;z_i)\prod_{j\neq i}V_{m_j,j_j}^{w_j}(x_j;z_j) \Bigr\rangle\\
&\hspace{2.5cm}+a_i^\Gamma\left(z-z_i\right)^{\frac{w_i-1}{2}}\Bigl\langle \prod_{\alpha=1}^{n+2g-2}W(u_{\alpha})\, V_{m_i+\frac{1}{2},j_i-\frac{1}{2}}^{w_i}(x_i;z_i)\prod_{j\neq i}V_{m_j,j_j}^{w_j}(x_j;z_j) \Bigr\rangle+\cdots\ ,
\end{split}
\end{equation}

\noindent where the $\cdots$ represent terms of order $(z-z_i)^{\frac{w_i+1}{2}}$. Now, by \eqref{eq:incidence-relation}, this expression vanishes exactly. Thus, comparing terms of lowest order, we find

\begin{equation}
\begin{split}
\Bigl\langle \prod_{\alpha=1}^{n+2g-2}W(u_{\alpha})\, &V_{m_i-\frac{1}{2},j_i+\frac{1}{2}}^{w_i}(x_i;z_i)\prod_{j\neq i}V_{m_j,j_j}^{w_j}(x_j;z_j) \Bigr\rangle\\
&=a_i^{\Gamma}\Bigl\langle \prod_{\alpha=1}^{n+2g-2}W(u_{\alpha})\, V_{m_i+\frac{1}{2},j_i+\frac{1}{2}}^{w_i}(x_i;z_i)\prod_{j\neq i}V_{m_j,j_j}^{w_j}(x_j;z_j) \Bigr\rangle\ .
\end{split}
\end{equation}

\noindent Shifting $j_i\to j_i+\frac{1}{2}$ and $m_i\to m_i+\frac{1}{2}$ gives us a recursion relation in the variable $m_i$ which is easily solvable. The general solution is given by

\begin{equation}\label{eq:recursion-solution}
\Bigl\langle\prod_{\alpha=1}^{n+2g-2}W(u_{\alpha})\prod_{i=1}^{n}V_{m_i,j_i}^{w_i}(x_i;z_i)\Bigr\rangle=C_{\Gamma}\prod_{i=1}^{n}(a_i^{\Gamma})^{-h_i}\ ,
\end{equation}

\noindent where $h_i=m_i+\frac{w_i}{2}$, and $C_{\Gamma}$ is a function of $j_i$, $x_i$, $z_i$, and $w_i$, but not of $m_i$.\footnote{Strictly speaking, $C_{\Gamma}$ can depend on the value of $m_i$ modulo $1/2$. However, this is no problem, since we are considering vertex operators which lie in irreducible representations of the symplectic Boson algebra \eqref{eq:symplectic-Boson-algebra}, for which $m_i\text{ mod }1/2$ is fixed.}

Now, equation \eqref{eq:recursion-solution} is valid when the configuration $\{z_i,x_i,w_i\}$ admits the covering map $\Gamma$. But, generically, as we have discussed, there are multiple points on the moduli space corresponding to separate covering maps. Thus, one could expect the form of the correlators as a function of $x_i,z_i$ to be a sum of distributions localizing on covering maps, multiplied by terms like the right-hand-side of \eqref{eq:recursion-solution}. If one assumes that these distributions are delta functions (see Section \ref{subsec:delta function}), we have

\begin{equation}\label{eq:general-recursion-solution}
\Bigl\langle\prod_{\alpha=1}^{n+2g-2}W(u_{\alpha})\prod_{i=1}^{n}V_{m_i,j_i}^{w_i}(x_i;z_i)\Bigr\rangle=\sum_{\Gamma}C_{\Gamma}\prod_{i=1}^{n}(a_i^{\Gamma})^{-h_i}\prod_{\ell=1}^{n+3g-3}\delta(f_{\ell}^{\Gamma})\ ,
\end{equation}

\noindent where $f_{\ell}^{\Gamma}$ are the $n+3g-3$ constraints for a covering map to exist. This is precisely the form of the correlators found in the RNS formalism in \cite{Eberhardt:2019,Eberhardt:2020}. However, unlike in \cite{Eberhardt:2019,Eberhardt:2020}, we have shown that \eqref{eq:general-recursion-solution} is the unique solution to the constraints imposed on our worldsheet correlators (if one assumes that the localizing distributions are delta functions). 

After integrating over the worldsheet moduli space $\mathcal{M}_{g,n}$, we have

\begin{equation}\label{eq:integrated-correlators}
\int_{\mathcal{M}_{g,n}}\mathrm{d}\mu\,\Bigl\langle\prod_{\alpha=1}^{n+2g-2}W(u_{\alpha})\prod_{i=1}^{n}V_{m_i,j_i}^{w_i}(x_i;z_i)\Bigr\rangle=\sum_{\Gamma}W_{\Gamma}\prod_{i=1}^{n}(a_i^{\Gamma})^{-h_i}\ .
\end{equation}

\noindent The appearance of the combination $\prod_{i}(a_i^{\Gamma})^{-h_i}$ in the correlation function is characteristic of correlation functions of twisted-sector ground states in the symmetric product orbifold, at least at genus zero \cite{Lunin_2002,Pakman_2009,Dei:2019}. Unfortunately, the general form of the contributions of higher genus covering spaces to correlation functions in the symmetric orbifold is not yet known. If, however, it were shown that the higher genus contributions were also proportional to the combination $\prod_{i}(a_i^{\Gamma})^{-h_i}$, then the integrated correlation function \eqref{eq:integrated-correlators} would exactly recover the behavior of the symmetric orbifold from $\rm{AdS}_3$ string theory, up to the prefactors $W_{\Gamma}$.\footnote{In principle, these prefactors can be fixed by the Knizhnik-Zamolodchikov equations on the worldsheet, but this analysis is quite complicated even at tree-level.}

\section{The covering map from local Ward identities}\label{sec:Ward-to-covering}

Equation \eqref{eq:incidence-relation} relates correlators with $\xi^{\pm}(z)$ insertions to the covering map. In \cite{Dei:2020} it was shown that, after multiplying by a simple prefactor, the correlation functions with $\xi^{-}(z)$ inserted could be interpreted as the numerator of $\Gamma$, while those with $\xi^+(z)$ inserted could be interpreted as the denominator of $\Gamma$. In this language, the local Ward identities\footnote{Here and throughout the paper, the term 'Ward identity' is loosely used to refer to constraints on correlation functions imposed by the symmetry algebra of the theory. The term 'local Ward identity' is used in the spirit of \cite{Dei:2020,Eberhardt:2019,Eberhardt:2020} to mean constraints on correlators imposed by the local behavior of OPEs of vertex operators with current fields.} constraining the correlation functions were shown to be manifestly identical to the constraints one would solve to construct a branched covering of the sphere. Here, we show how this analysis extends to higher genus Riemann surfaces. We do this by showing that the set of algebraic identities constraining the covering map are exactly identical to the local Ward identities constraining the correlators, coming from the OPEs of $\xi^{\pm}$ with the vertex operators. The equivalence of the constraints on the covering map with the Ward identities on the string theory correlators comprises a second, more constructive, proof of the relation \eqref{eq:incidence-relation}, closer in spirit to that given in \cite{Dei:2020}.

\subsection{Constraints on the covering map}

We begin by writing an algebraic set of equations which can be solved to determine the covering map $\Gamma$ for a given set of points $z_i$ on $\Sigma$, points $x_i$ on $\rm{S}^2$, and branching indices $w_i$.

Let $\vartheta(x,y)$ denote the prime form on $\Sigma$, the definition of which is reviewed in Section \ref{subsec:prime-form}. Since $\Gamma:\Sigma\to\rm{S}^2$ is a holomorphic map of order $N$, given by the Riemann-Hurwitz formula \eqref{eq:Riemann-Hurwitz}, its divisor $(\Gamma)$ can be written as

\begin{equation}
(\Gamma)=\sum_{a=1}^{N}Q_a-\sum_{a=1}^{N}P_a\ ,
\end{equation}

\noindent where $Q_a,P_a$ are points on $\Sigma$, namely the zeroes and poles of $\Gamma$, respectively (for an introduction to divisors, see Section \ref{subsec:divisors}). Since $\Gamma$ is a globally-defined meromorphic function on $\Sigma$, Abel's theorem tells us that the Abel-Jacobi map $\mu:\text{Div}(\Sigma)\to\text{Jac}(\Sigma)$ must send $(\Gamma)$ to the origin (see Section \ref{subsec:Abel}). That is, the poles and zeroes of $\Gamma$ must satisfy

\begin{equation}\label{eq:abel-theorem-covering}
\sum_{a=1}^{N}\mu(Q_a)\equiv\sum_{a=1}^{N}\mu(P_a)\quad\text{ mod }\mathbb{Z}^g+\Omega\mathbb{Z}^g\ .
\end{equation}

\noindent This is the first set of constraints imposed on the covering map.

For the second, we use the fact that, assuming we know the zeroes and poles of $\Gamma$, we can express it uniquely as

\begin{equation}
\Gamma(z)=C\left(\prod_{a=1}^{N}\vartheta(z,Q_a)\right)\left(\prod_{a=1}^{N}\vartheta(z,P_a)\right)^{-1}\ ,
\end{equation}

\noindent where $C$ is some multiplicative constant. Now, we write

\begin{equation}
P^+(z)=C^+\prod_{a=1}^{N}\vartheta(z,P_a)\ ,\quad P^-(z)=C^-\prod_{a=1}^{N}\vartheta(z,Q_a)\ ,
\end{equation}

\noindent where $C^{\pm}$ are constants such that $-C^-/C^+=C$ (we choose two separate constants for similarity to the constraints arising from the Ward identities in the next subsection). Both $P^{\pm}(z)$ are individually non-periodic, but have the same monodromies around the cycles of $\Sigma$ by \eqref{eq:abel-theorem-covering}, so that the ratio $P^{-}/P^{+}=-\Gamma(z)$ is well-defined on $\Sigma$. Now, the only requirement for $\Gamma$ to be a covering map is that $\Gamma$ maps $z_i$ to $x_i$ and has a critical point of order $w_i$ at $x_i$. Algebraically,

\begin{equation}
\Gamma(z)-x_i\sim\mathcal{O}\left((z-z_i)^{w_i}\right)\ ,\quad z\to z_i\ .
\end{equation}

\noindent Phrased in terms of the numerator and denominator $P^-$ and $P^+$, these constraints take the form

\begin{equation}\label{eq:covering-constraint}
P^-(z)+x_iP^+(z)\sim\mathcal{O}\left((z-z_i)^{w_i}\right)\ ,\quad z\to z_i\ .
\end{equation}

Equations \eqref{eq:abel-theorem-covering} and \eqref{eq:covering-constraint} give us $\sum_{i=1}^{n}w_i+g=2N+n-2+3g$ constraints for the $2N+1$ unknowns $P_a,Q_a,C$. Thus, the existence of a solution requires the fine-tuning of $n+3g-3=\text{dim}(\mathcal{M}_{g,n})$ moduli on the worldsheet, which agrees with the statement in Section \ref{sec:Incidence-relation} that the covering map exists on discrete points on the moduli space $\mathcal{M}_{g,n}$.

\subsection{The local Ward identities}

Now we shall derive a set of constraints on the correlation functions for the correlation functions imposed by the local structure of correlation functions with $\xi^{\pm}$ insertions, which we call 'local Ward identities' in the spirit of \cite{Eberhardt:2019,Eberhardt:2020,Dei:2020}. To do this, we investigate the analytic structure of the functions

\begin{equation}
\omega^{\pm}(z)\equiv\Bigl\langle\xi^{\pm}(z)\prod_{\alpha=1}^{n+2g-2}W(u_{\alpha})\prod_{i=1}^{n}V_{m_i,j_i}^{w_i}(x_i;z_i)\Bigr\rangle\ .
\end{equation}

\noindent As we remarked in Section \ref{sec:Incidence-relation}, these functions define $1/2$-forms on $\Sigma$. Thus, the divisors $(\omega^{\pm})$ have degree $g-1$, and we will denote them by $D^{\pm}$. Every pole of $\omega^{\pm}$ comes from the OPEs of $\xi^{\pm}$ with $V_{m_i,j_i}^{w_i}$, and we also know that $D^{\pm}$ contains simple zeroes at $z=u_{\alpha}$, due to the regularity of the $\xi^{\pm}W$ OPE. Thus, there exists positive divisors $\Delta^{\pm}$ such that

\begin{equation}\label{splitting-divisors}
D^{\pm}=-\sum_{i=1}^{n}\frac{w_i+1}{2}z_i+\sum_{\alpha=1}^{n+2g-2}u_{\alpha}+\Delta^{\pm}.
\end{equation}

\noindent The degrees of the divisors $\Delta^{\pm}$ are easily calculated to be

\begin{equation}
\deg{\Delta^{\pm}}=\deg D^{\pm}+\sum_{i=1}^{n}\frac{w_i+1}{2}-n-2g+2=N.
\end{equation}

\noindent That is, $\omega^{\pm}(z)$ have $N$ ``extra'' zeroes that we have not yet accounted for. Let us denote them explicitly as

\begin{equation}
\Delta^+=\sum_{a=1}^{N}P_a\ ,\quad\quad\Delta^-=\sum_{a=1}^{N}Q_a\ .
\end{equation}

Now, let us consider the ``rescaled'' forms

\begin{equation}
\begin{split}
\widetilde{\omega}^+(z)&=\prod_{i=1}^{n}\vartheta(z,z_i)^{\frac{w_i+1}{2}}\prod_{\alpha=1}^{n+2g-2}\vartheta(z,u_{\alpha})^{-1}\prod_{a=1}^{N}\vartheta(z,P_a)^{-1}\,\,\omega^{+}(z)\ ,\\
\widetilde{\omega}^-(z)&=\prod_{i=1}^{n}\vartheta(z,z_i)^{\frac{w_i+1}{2}}\prod_{\alpha=1}^{n+2g-2}\vartheta(z,u_{\alpha})^{-1}\prod_{a=1}^{N}\vartheta(z,Q_a)^{-1}\,\,\omega^{-}(z)\ .
\end{split}
\end{equation}

\noindent Since $\vartheta(z,a)$ defines a $-1/2$-form in $z$, $\widetilde{\omega}^{\pm}$ both behave as meromorphic $g/2$-forms. Furthermore, $\widetilde{\omega}^{\pm}$ have no zeroes or poles. Finally, they are not globally defined on $\Sigma$, but instead pick up a phase around the $\beta$-cycles, given by

\begin{equation}
\widetilde{\omega}^{\pm}(z+\beta_{\mu})=\exp\left(-i\pi(g-1)\Omega_{\mu\mu}+2\pi i\int_{(g-1)z}^{\Delta}\omega_{\mu}\right)\,\widetilde{\omega}^{\pm}(z)\ ,
\end{equation}

\noindent where $\Delta$ is a divisor such that $2\Delta$ is in the canonical class, and such that $\Delta$ is in the same divisor class as $D^{\pm}$, and $\omega_{\mu}(z)$ are the canonical basis of holomorphic 1-forms on $\Sigma$ (see \cite{Verlinde:1987,Hikida_2020}).\footnote{That is, $\Delta$ is the divisor class corresponding to the chosen spin structure on $\Sigma$. The statement that $\Delta$ is in the same divisor class as $D^{\pm}$ is simply the statement that we choose the spin structure on $\Sigma$ to be determined by the divisors $D^{\pm}$ of $\omega^{\pm}$.} These properties determine $\widetilde{\omega}^{\pm}(z)$ up to a constant \cite{Verlinde:1987}. That is, there is a unique (up to an overall factor) $g/2$-form $\sigma(z)$ such that

\begin{equation}\label{eq:sigma}
\widetilde{\omega}^{\pm}(z)=C^{\pm}\sigma(z)
\end{equation}

\noindent for some constants $C^{\pm}$. We prove that such a $g/2$-form $\sigma$ exists and is unique in Appendix \ref{appendix:sigma}. Using the definitions of $\widetilde{\omega}^{\pm}(z)$, we thus conclude that

\begin{equation}\label{eq:omega-pm-ansatz}
\begin{split}
\omega^+(z)&=C^+\prod_{i=1}^{n}\vartheta(z,z_i)^{-\frac{w_i+1}{2}}\prod_{\alpha=1}^{n+2g-2}\vartheta(z,u_{\alpha})\prod_{a=1}^{N}\vartheta(z,P_a)\,\,\sigma(z)\ ,\\
\omega^-(z)&=C^-\prod_{i=1}^{n}\vartheta(z,z_i)^{-\frac{w_i+1}{2}}\prod_{\alpha=1}^{n+2g-2}\vartheta(z,u_{\alpha})\prod_{a=1}^{N}\vartheta(z,Q_a)\,\,\sigma(z)\ .
\end{split}
\end{equation}

\noindent That is, we can completely parametrize $\omega^{\pm}(z)$ in terms of the $2N$ points $P_a,Q_a$ and two unknown constants $C^{\pm}$.

The above parametrization of $\omega^{\pm}(z)$ can be naturally related to the branched covering map $\Gamma$. To see this, define the quasi-periodic functions

\begin{equation}\label{eq:P-pm-def}
\begin{split}
P^+(z)&:=\prod_{i=1}^{n}\vartheta(z,z_i)^{\frac{w_i+1}{2}}\prod_{\alpha=1}^{n+2g-2}\vartheta(z,u_{\alpha})^{-1}\frac{\omega^+(z)}{\sigma(z)}=C^+\prod_{a=1}^{N}\vartheta(z,P_a)\ ,\\
P^-(z)&:=\prod_{i=1}^{n}\vartheta(z,z_i)^{\frac{w_i+1}{2}}\prod_{\alpha=1}^{n+2g-2}\vartheta(z,u_{\alpha})^{-1}\frac{\omega^+(z)}{\sigma(z)}=C^-\prod_{a=1}^{N}\vartheta(z,Q_a)\ .
\end{split}
\end{equation}

\noindent Now, we can rewrite the identities constraining $\omega^{\pm}(z)$ into constraints on the leftover variables $P_a,Q_a$ and $C^{\pm}$. To begin, by \eqref{eq:regular-OPE}, one finds that

\begin{equation}\label{eq:correlators-constraint}
P^-(z)+x_kP^+(z)\sim\mathcal{O}((z-z_k)^{w_k}),\quad z\to z_k\ .
\end{equation}

\noindent Furthermore, since $\widetilde{\omega}^{+}$ and $\widetilde{\omega}^{-}$ have the same branch points ($z=z_i$ with $w_i$ even) and are defined with respect to the same spin structure, their ratio $\omega^-(z)/\omega^+(z)$ formally defines a meromorphic function whose divisor is $D^+-D^-$. By Abel's theorem, we must have $\mu(D^+)=\mu(D^-)$ in order for such a meromorphic function to exist. Using \eqref{splitting-divisors}, this gives the requirement that

\begin{equation}\label{eq:abel-theorem-correlators}
\sum_{a=1}^{N}\mu(Q_a)=\sum_{a=1}^{N}\mu(P_a)\ .
\end{equation}

\noindent Equations \eqref{eq:correlators-constraint} and \eqref{eq:abel-theorem-correlators} are \textit{precisely} the algebraic conditions \eqref{eq:covering-constraint} and \eqref{eq:abel-theorem-covering} that define a branched covering $\Gamma:\Sigma\to\rm{S}^2$. In particular, we can interpret $P_a$ and $Q_a$ as the locations of the poles and zeroes of the covering map, respectively. This gives the interpretation of $P^+(z)$ and $P^-(z)$ as the denominator and numerator of the covering map, and the incidence relation $P^-(z)+\Gamma(z)P^+(z)=0$ is manifest. In \cite{Dei:2020}, this argument was the basis of the proof of the genus zero analogue of the incidence relation \eqref{eq:incidence-relation}, and thus this section provides a natural extension of that result at higher genus.

\subsection{Delta function localization}\label{subsec:delta function}

We have shown that we can recast the Ward identities for the correlators $\omega^{\pm}(z)$ as a system of $\sum_{i}w_i+g=2N+n+3g-2$ equations in $2N+2$ variables, which are homogeneous in the sense that $C^{\pm}$ can be determined only up to a common scaling factor. The existence of a solution to such a system requires the fine-tuning of $n+3g-3$ variables. If we treat $x_i$ as fixed, this requires all $n+3g-3$ moduli on the worldsheet $\Sigma$ to be fine-tuned.

However, if we allow the constants $C^{\pm}$ to be distributions on the moduli space, then we can come up with a solution to \eqref{eq:correlators-constraint} and \eqref{eq:abel-theorem-correlators} which is valid at all points in the moduli space, not just those for which a covering map exists. In this case, we would write our ansatz \eqref{eq:omega-pm-ansatz} as a sum over all covering maps allowed for a given $\{x_i,w_i\}$, weighted by an appropriate distribution $C^{\pm}_{\Gamma}$. In terms of the variables $P^{\pm}(z)$, we rewrite \eqref{eq:P-pm-def} as

\begin{equation}
P^+(z)=\sum_{\Gamma}C^+_{\Gamma}\prod_{a=1}^{N}\vartheta(z,P_a^{\Gamma})\ ,\quad P^-(z)=\sum_{\Gamma}C^-_{\Gamma}\prod_{a=1}^{N}\vartheta(z,Q_a^{\Gamma})\ .
\end{equation}

\noindent In \cite{Dei:2020} it was shown that at genus zero, the distributions $C^{\pm}_{\Gamma}$ are delta functions on the moduli space. It seems natural, then, that this would also be the case at higher genus. However, unlike in \cite{Dei:2020}, the precise distributional nature of $C^{\pm}_{\Gamma}$ is quite difficult to pin down, mostly due to the nonlinear dependence of the prime form on the insertion point $z$ and the moduli of $\Sigma$. However, it is simple to show that, if one assumes that $C^{\pm}_{\Gamma}$ is a delta function, then equation \eqref{eq:correlators-constraint} is satisfied, provided that $P_a^{\Gamma}$ and $Q_a^{\Gamma}$ are chosen as the poles and zeroes of the covering map in question. To see this, we take

\begin{equation}
\begin{split}
P^+(z)=\sum_{\Gamma}K^+_{\Gamma}\prod_{\ell=1}^{n+3g-3}\delta(f_\ell^{\Gamma})\prod_{a=1}^{N}\vartheta(z,P_a^{\Gamma})\ ,\\
P^-(z)=\sum_{\Gamma}K^-_{\Gamma}\prod_{\ell=1}^{n+3g-3}\delta(f_\ell^{\Gamma})\prod_{a=1}^{N}\vartheta(z,Q_a^{\Gamma})\ ,
\end{split}
\end{equation}

\noindent where $f_\ell^\Gamma$ are a set of $n+3g-3$ conditions on the moduli for the covering map to exist and $K^{\pm}_{\Gamma}$ are some normalization constants. Plugging this ansatz into \eqref{eq:correlators-constraint} then yields

\begin{equation}
P^-(z)+x_kP^+(z)=\sum_{\Gamma}\left(K^-_{\Gamma}\prod_{a=1}^{N}\vartheta(z,Q_a^{\Gamma})+x_kK^+_{\Gamma}\prod_{a=1}^{N}\vartheta(z,P_a^{\Gamma})\right)\bigg|_{f_{\ell}^{\Gamma}=0}\prod_{\ell=1}^{n+3g-3}\delta(f_{\ell}^{\Gamma})\ .
\end{equation}

\noindent However, since we have chosen $Q_a^{\Gamma}$ and $P_a^{\Gamma}$ to be the zeroes and poles of the covering map $\Gamma$, when $f_{\ell}^{\Gamma}=0$, the term in parentheses is exactly the numerator of $\Gamma(z)-x_k$. Thus, if we take $z\to z_k^{\Gamma}$, we have

\begin{equation}
P^-(z)+x_kP^+(z)\sim\mathcal{O}((z-z_k^{\Gamma})^{w_k})\ ,\quad z\to z_k^{\Gamma}\ ,
\end{equation}

\noindent where $z_k^{\Gamma}=\Gamma^{-1}(x_k)$. Therefore, the delta function ansatz satisfies equation \eqref{eq:correlators-constraint} with $z_i=z_i^{\Gamma}$.

\section{Discussion}\label{sec:discussion}

We have shown that the class of correlation functions \eqref{eq:generic-correlator} in the $\mathfrak{sl}(2,\mathbb{R})_1\oplus\widehat{\mathfrak{u}(1)}$ WZW model naturally reproduce correlation functions of twisted-sector ground states in the symmetric orbifold theory at every order in string perturbation theory. Since these are the correlation functions describing the scattering of spectrally-flowed highest-weight states in string theory on $\rm{AdS}_3\times\rm{S}^3\times\mathbb{T}^4$, this result represents highly nontrivial evidence that the worldsheet theory is exactly dual to the symmetric orbifold $\rm{Sym}^{N}(\mathbb{T}^4)$. Furthermore, this provides a precise realization of the conjecture put forth in \cite{Pakman_2009} that the string worldsheet should be identified with the covering space used in the construction of symmetric orbifold correlation functions \cite{Lunin_2001}.

This work is essentially a generalization of the results of \cite{Dei:2020}, for which the identification of the string worldsheet with the covering surface in the dual CFT was done for lowest-order (spherical) contributions. By demonstrating that this analysis can be generalized to include each order in string perturbation theory, we have shown that the free field realization employed here and in \cite{Dei:2020} is an incredibly powerful tool that is bound to be useful in establishing this incarnation of the holographic correspondence further.

\vspace{0.5cm}

\noindent We conclude with a list of open questions that could potentially be tackled with the formalism developed here:


\subsection*{\boldmath Nontrivial $\rm{AdS}_3$ boundaries}

So far, we have only considered the worldsheet theory of the tensionless string on global $\rm{AdS}_3$, for which the boundary has a spherical topology. However, there are local $\rm{AdS}_3$ spacetimes which admit topologically nontrivial boundaries, for which the spectrum of the tensionless limit of string theory has been recently computed \cite{eberhardt2020partition}. It would be interesting to see if our free field analysis could be applied to these spacetimes. In this case, it would be natural to expect one to be able to write the correlation functions in terms of covering maps $\Gamma:\Sigma\to\partial\rm{AdS}_3$ from the worldsheet to the nontrivial $\rm{AdS}_3$ boundary. Such an identification would naturally relate the worldsheet theory to the symmetric orbifold CFT defined on the Riemann surface $\partial\rm{AdS}_3$.

\subsection*{Nonperturbative effects}

This paper, as a natural extension of \cite{Dei:2020}, demonstrates the equivalence between correlation functions in tensionless $\rm{AdS}_3\times\rm{S}^3\times\mathbb{T}^4$ string theory to those of the symmetric product CFT $\text{Sym}^N(\mathbb{T}^4)$ to all orders in the $1/N$ expansion (equivalently, to all orders in string perturbation theory). However, string theory famously is riddled with nonperturbative excitations (branes) that contribute beyond any order in perturbation theory. Thus, in order to truly establish a holographic dictionary between tensionless $\rm{AdS}_3$ string theory and the dual symmetric product CFT, it is necessary to study these nonperturbative effects. On the string side, these effects would be in the form of D-branes (or $\rm{AdS}_2$ branes \cite{Bachas:2002-perm,Bachas:2002-asymp}). However, as was mentioned in Section \ref{sec:Incidence-relation}, the perturbation series for tensionless $\rm{AdS}_3$ string theory naively truncates. Since the $1/N$ expansion in the symmetric product CFT is exact at finite $N$ \cite{Lunin_2002}, one would naively expect such nonperturbative effects to be absent if tensionless $\rm{AdS}_3\times\rm{S}^3\times\mathbb{T}^4$ string theory is to really be dual to $\text{Sym}^N(\mathbb{T}^4)$. It would thus be natural to suggest that, in the presence of such branes, the dual boundary CFT may be modified in such a way that a holographic dual to D-branes becomes present.

\subsection*{\boldmath Strings on $\rm{AdS}_3\times\rm{S}^3\times\rm{S}^3\times\rm{S}^1$}

This work utilized a free field realization of string theory on $\rm{AdS}_3\times\rm{S}^3\times\mathbb{T}^4$ to constrain correlation functions in such a way that they are manifestly related to correlation functions in the symmetric product CFT $\rm{Sym}^N(\mathbb{T}^4)$. However, if the NS-NS flux through the two $\text{S}^3$'s agree, then string theory on $\rm{AdS}_3\times\rm{S}^3\times\rm{S}^3\times\rm{S}^1$ in the hybrid formalism also admits a free field realization in terms of the superalgebra $\mathfrak{d}(2,1;\alpha)$ with $\alpha=1$ \cite{Eberhardt:2019niq}, and is conjectured to be dual to a symmetric product CFT as well \cite{Eberhardt:2017fsi,Eberhardt:2017pty}. Thus, one would expect a similar analysis to be valid for strings on this spacetime. However, since the free-field realization of $\mathfrak{d}(2,1;\alpha)$ does not require one to take any cosets (as in the $\mathfrak{psu}(1,1|2)_1$ model), there are no fields in the worldsheet formalism on $\rm{AdS}_3\times\rm{S}^3\times\rm{S}^3\times\rm{S}^1$ analogous to the $W(u)$ operators that we so heavily employed in this paper. Thus, while it is certainly possible that an identity analogous to \eqref{eq:incidence-relation} applies to the case of strings on $\rm{AdS}_3\times\rm{S}^3\times\rm{S}^3\times\rm{S}^1$ as well, its proof is bound to be a bit different, and perhaps more subtle.

\section*{Acknowledgements}

I would like to thank Matthias Gaberdiel, Rajesh Gopakumar and Andrea Dei for their roles in starting the project that resulted in this paper. I thank Matthias Gaberdiel and Andrea Dei for helpful comments on an early version of the manuscript, and Lorenz Eberhardt for informative and enlightening discussions. Finally, I would like to thank my parents, Robert and Sherry Knighton, for their hospitality during the final stages of this work. This work was supported by the Swiss National Science Foundation through a personal grant and via the NCCR SwissMAP.

\newpage

\appendix

\section{Some Riemann surface theory}\label{sec:Riemann-Surfaces}

In this appendix, we introduce some of the main results of the study of compact Riemann surfaces. In particular, we introduce meromorphic functions and forms, divisors, spin structures, Abel's theorem, and the construction of meromorphic functions from their divisors. We will assume that the reader has a basic understanding of Riemann surfaces as one dimensional complex manifolds. For an excellent introduction to the theory of Riemann surfaces, see \cite{donaldson_2011}. For the most part, we follow the notation and conventions of \cite{eynard2018lectures}.

\begin{figure}[!ht]
\centering
\begin{tikzpicture}
\draw[smooth] (0,1) to[out=30,in=150] (2,1) to[out=-30,in=210] (3,1) to[out=30,in=150] (5,1) to[out=-30,in=30] (5,-1) to[out=210,in=-30] (3,-1) to[out=150,in=30] (2,-1) to[out=210,in=-30] (0,-1) to[out=150,in=-150] (0,1);
\draw[smooth] (0.4,0.1) .. controls (0.8,-0.25) and (1.2,-0.25) .. (1.6,0.1);
\draw[smooth] (0.5,0) .. controls (0.8,0.2) and (1.2,0.2) .. (1.5,0);
\draw[smooth] (3.4,0.1) .. controls (3.8,-0.25) and (4.2,-0.25) .. (4.6,0.1);
\draw[smooth] (3.5,0) .. controls (3.8,0.2) and (4.2,0.2) .. (4.5,0);
\draw[blue] (-0.5,0) arc(180:0:0.51 and 0.2);
\draw[blue, dashed] (-0.5,0) arc(180:0:0.51 and -0.2);
\draw[blue] (5.5,0) arc(180:0:-0.51 and 0.2);
\draw[blue, dashed] (5.5,0) arc(180:0:-0.51 and -0.2);
\draw[red] (1,0.1) ellipse (1 and 0.6);
\node[above] at (1,0.6) {$\beta_1$};
\draw[red] (4,0.1) ellipse (1 and 0.6);
\node[above] at (4,0.6) {$\beta_2$};
\draw[white] (-1,-.25) -- (1,-2.5);
\end{tikzpicture}
\hspace{2cm}
\begin{tikzpicture}[scale = 0.75]
\begin{scope}[very thick,decoration={markings,mark=at position 0.5 with {\arrow{>}}}]
	\draw[red, postaction = {decorate}] (-1,2.44) -- (1,2.44);
	\draw[blue, postaction = {decorate}] (1,2.44) -- (2.44,1);
	\draw[red, postaction = {decorate}] (2.44,-1) -- (2.44,1);
	\draw[blue, postaction = {decorate}] (1,-2.44) -- (2.44,-1);
	\draw[red, postaction = {decorate}] (1,-2.44) -- (-1,-2.44);
	\draw[blue, postaction = {decorate}] (-1,-2.44) -- (-2.44,-1);
	\draw[red, postaction = {decorate}] (-2.44,1) -- (-2.44,-1);
	\draw[blue, postaction = {decorate}] (-1,2.44) -- (-2.44,1);
\end{scope}
\node[above left] at (-1.72,1.72) {\large$\alpha_1$};
\node[above right] at (1.72,1.62) {\large$\alpha_1$};
\node[above] at (0,2.5) {\large$\beta_1$};
\node[right] at (2.5,0) {\large$\beta_1$};
\node[below right] at (1.72,-1.72) {\large$\alpha_2$};
\node[below left] at (-1.72,-1.72) {\large$\alpha_2$};
\node[below] at (0,-2.5) {\large$\beta_2$};
\node[left] at (-2.5,0) {\large$\beta_2$};
\end{tikzpicture}
\caption{Left: A genus 2 surface with canonical homology basis $\alpha_1,\alpha_2,\beta_1,\beta_2$. Right: The same surface modelled as an $4g=8$-gon with identified edges representing the $\alpha$ and $\beta$ cycles.}
\label{fig:Riemann-surface}
\end{figure}
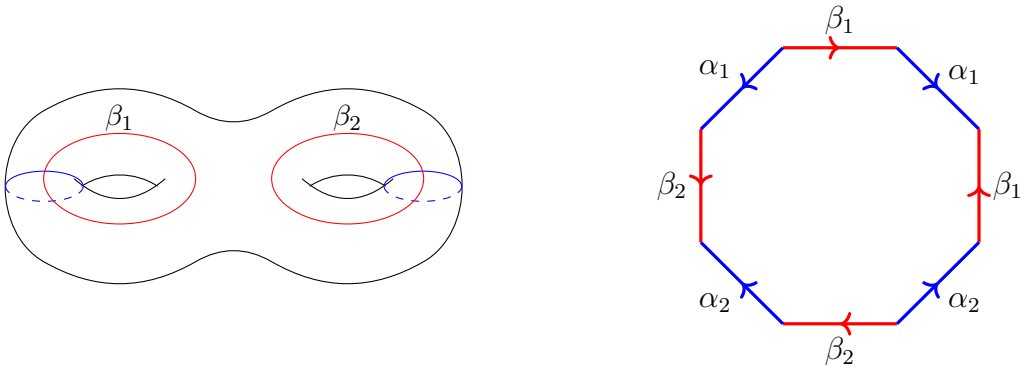

Before diving into the theory, let us first fix our conventions. Consider a genus $g$ Riemann surface $\Sigma$. We can choose a basis for the homology group $H_1(\Sigma,\mathbb{Z})$ as in Figure \ref{fig:Riemann-surface}, consisting of cycles $\alpha_{\mu},\beta_{\mu}$ with the intersection numbers

\begin{equation}
\alpha_{\mu}\cap\alpha_{\nu}=0\ ,\quad \beta_{\mu}\cap\beta_{\nu}=0\ ,\quad \alpha_{\mu}\cap\beta_{\nu}=\delta_{\mu\nu}\ .
\end{equation}

\noindent This choice of homology basis is essentially unique, up to a $\text{Sp}(2g,\mathbb{Z})$ transformation which leaves these intersections invariant. All of the analysis we will do in this text is independent of the particular basis $\alpha_{\mu},\beta_{\mu}$, but we fix one for concreteness. A Riemann surface $\Sigma$ together with a choice of homology basis $\alpha_{\mu},\beta_{\mu}$ is called a \textit{marked} Riemann surface. A marked Riemann surface $\Sigma$ is moreover homeomorphic to a $4g$-gon whose edges are $\alpha_{\mu}$ and $\beta_{\mu}$, glued together as in Figure \ref{fig:Riemann-surface}.

Furthermore, the surface $\Sigma$ admits a basis of $g$ linearly independent holomorphic one-forms $\omega_1,\ldots,\omega_{g}$. One can choose this basis such that

\begin{equation}
\oint_{\alpha_{\mu}}\omega_{\nu}=\delta_{\mu\nu}\ .
\end{equation}

\noindent That is, such that $\omega_{\mu}$ is dual to $\alpha_{\mu}$. Once such a basis is picked, the integral of $\omega_{\nu}$ over the cycle $\beta_{\mu}$ is left unconstrained. We thus define a complex $g\times g$ matrix $\Omega$, the so-called \textit{period matrix}, via

\begin{equation}
\Omega_{\mu\nu}=\oint_{\beta_{\mu}}\omega_{\nu}\ .
\end{equation}

\noindent A fundamental result in the theory of Riemann surfaces, the Riemann bilinear relations, implies that $\Omega$ is a symmetric matrix and that $\text{Im}(\Omega)$ is a positive-definite matrix.

\subsection{Meromorphic forms}\label{subsec:forms}

With our basic notions out of the way, we can now introduce the primary object of study -- meromorphic forms on the Riemann surface $\Sigma$.

On a generic manifold, a differential 1-form is a linear combination $A_{\mu}(x)\mathrm{d}x^{\mu}$ of the basis vectors $\mathrm{d}x^{\mu}$ with coefficients being functions on the manifold. On a Riemann surface, the appropriate definition would be either $f(z,\overline{z})\,\mathrm{d}z$, $f(z,\overline{z})\,\mathrm{d}\overline{z}$, or some linear combination thereof. A \textit{meromorphic form of weight 1} on the surface $\Sigma$ is a form which can be written locally as $f(z)\,\mathrm{d}z$ for some meromorphic function $f(z)$. Similarly, an anti-meromorphic form of weight 1 would be expressed as $f(\overline{z})\,\mathrm{d}\overline{z}$. In geometric terms, one can consider a meromorphic form to be a section of a certain line bundle over $\Sigma$, which we will call $K$ (often called the \textit{canonical line bundle}).\footnote{We will later also use the letter $K$ for the divisor class of this bundle. It is unfortunately quite common to denote line bundles and divisor classes with the same letters.} The space of sections $\Gamma(K)$ on $K$ is often written as $\mathcal{M}^1(\Sigma)$. That is, $\mathcal{M}^1(\Sigma)$ is the space of meromorphic forms of weight 1 on $\Sigma$.

We can construct meromorphic forms of other weights as well. In general, a meromorphic form of weight $h$ is an object locally expressible as $f(z)\,(\mathrm{d}z)^h$. Such an object is a section of the line bundle $K^h$, and we call the space of such forms $\mathcal{M}^h(\Sigma)$. The space $\mathcal{M}^0(\Sigma)$ is just the space of meromorphic functions, and will often be abbreviated as $\mathcal{M}(\Sigma)$ (without the superscript). We say a meromorphic form is \textit{holomorphic} if it is locally expressable as a holomorphic function $f(z)\,(\mathrm{d}z)^h$ -- i.e. if it has no poles. We denote the space of holomorphic forms of weight $h$ by $\mathcal{O}^h(\Sigma)$.

Now, more abstractly, one can specify a meromorphic form to be an object $\omega(z)$ which, under the conformal coordinate transformation $z\to\widetilde{z}$, transforms according to the rule

\begin{equation}
\omega(\widetilde{z})=\left(\frac{\mathrm{d}\widetilde{z}}{\mathrm{d}z}\right)^h\omega(z)\ .
\end{equation}

\noindent Thus, the relationship between meromorphic forms on $\Sigma$ and conformal field theory becomes quite evident -- the transformation law of a meromorphic form of weight $h$ is precisely the data of a chiral field of conformal weight $h$. That is,

\begin{equation}
\text{meromorphic forms of weight }h\Longleftrightarrow\text{conformal fields of weight }h\ .
\end{equation}

\noindent This relationship is precisely why the meromorphic forms are useful in the study of conformal field theory on Riemann surfaces.

\subsection{Divisors}\label{subsec:divisors}

Now that we have defined meromorphic forms, we turn our head to divisors. A \textit{divisor} is a formal (finite) linear sum of points on the surface $\Sigma$, with coefficients in $\mathbb{Z}$ (or $\mathbb{Z}/2$, for spinor forms). That is, given a collection of points $a_i\in\Sigma$, we define the divisor $D$ to be

\begin{equation}
D=\sum_{i}n_ia_i
\end{equation}

\noindent for some integers $n_i$. Given two divisors $D$ and $D'$, their sum is given by adding their coefficients (where the coefficient of a point is taken to be zero if it does not contribute to the sum). Thus, the set of divisors on $\Sigma$ form a free Abelian group $\text{Div}(\Sigma)$ generated by points on $\Sigma$. The degree $\text{deg}(D)$ of a divisor is defined to be the sum of its coefficients.

A divisor's place in life is to represent the set of zeroes and poles of some meromorphic function $f(z)$. That if, given a meromorphic function $f(z)$ with zeroes of order $n_i$ at points $a_i$ and poles of order $m_i$ at the points $b_i$, we can associate a divisor $(f)$ to $f(z)$ given by

\begin{equation}
(f)=\sum_{i}n_ia_i-\sum_{i}m_ib_i\ .
\end{equation}

\noindent That is, the divisor of $f$ is the set of its zeroes and poles, counting multiplicities, where a pole is treated as a zero of negative order. If $f$ has no poles, the divisor $(f)$ has all positive coefficients. We call a divisor with no negative coefficients a \textit{positive divisor}, for obvious reasons. 

An immediate consequence of the definition of the divisor of a meromorphic function is that the divisor of the function $f(z)g(z)$ is the sum of the divisors of $f(z)$ and $g(z)$, i.e.

\begin{equation}
(f\cdot g)=(f)+(g)\ .
\end{equation}

\noindent That is, the map $(\cdot):\mathcal{M}(\Sigma)\to\text{Div}(\Sigma)$ is a homomorphism of abelian groups.

We can also talk about divisors of meromorphic forms of weight $h$ in exactly the same way, by counting the zeroes and poles weighted by their multiplicity. That is, there is a map $(\cdot):\mathcal{M}^h(\Sigma)\to\text{Div}(\Sigma)$ in the same way as there is for meromorphic functions.

It is often too cumbersome to study the space of divisors itself, since it is an enormous Abelian group. A much smaller group can be generated by quotienting $\text{Div}(\Sigma)$ by the subset of divisors which are the vanishing sets of some meromorphic function. Specifically, we call a divisor $P\in\text{Div}(\Sigma)$ a \textit{principal divisor} if $P=(f)$ for some meromorphic function $f$. We define an equivalence relation $\sim$ by

\begin{equation}
D\sim D'\Longleftrightarrow D=D'+P\text{ for some principal divisor }P\ .
\end{equation}

\noindent The space $\text{Div}(\Sigma)/\sim$ is called the \textit{divisor class group}, and is typically denoted $\text{Cl}(\Sigma)$. One particularly nice feature of the divisor class group is that any two meromorphic forms of the same (integer) weight $h$ have the same divisor, up to equivalence. This is simple to see since, given meromorphic forms $f(z)\,(\mathrm{d}z)^h$ and $g(z)\,(\mathrm{d}z)^h$, their ratio is a well-defined meromorphic function. Thus, to each bundle $K^h$, we can associate a unique element of $\text{Cl}(\Sigma)$.\footnote{As it turns out, this correspondence is bijective. Specifically, the divisor class group is isomorphic to the \textit{Picard group} $\text{Pic}(\Sigma)$ of equivalence classes of holomorphic line bundles on $\Sigma$.} For this reason, we denote the divisor class of differential forms of weight 1 by $K$, the so-called \textit{canonical class} on $\Sigma$.

A central result is that the elements of the trivial divisor class (i.e. the divisors of meromorphic functions) have degree zero. Put plainly, if a meromorphic function has zeroes at $a_i$ of order $n_i$ and poles at $b_i$ of order $m_i$, then

\begin{equation}
\sum_{i}n_i=\sum_{i}m_i\ .
\end{equation}

\noindent Put another way, if $f$ is a meromorphic function, then $\text{deg}((f))=0$.\footnote{This is not true for non-compact Riemann surfaces. For example, on $\mathbb{C}$ the function $f(z)=z$ has one zero but no poles, and thus $\text{deg}((f))=1$.} As a consequence, the elements of any divisor class all have the same degree, since they differ by the divisor of a meromorphic function, whose degree vanishes. Since the divisor classes we care about are of the form $h\cdot K$ for some integer $h$, we know the degree of all meromorphic forms if we know the degree of those forms with weight $h=1$. This degree is dependent on the genus $g$ of the surface $\Sigma$, and is known to be

\begin{equation}
\text{deg}(K)=2g-2\ .
\end{equation}

\noindent Thus, since the degree of the sum of two divisors is just the sum of the degrees, the degree of a meromorphic form of weight $h$ is simply

\begin{equation}
\text{deg}(h\cdot K)=(2g-2)h\ .
\end{equation}

\subsection{Existence of forms and Abel's theorem}\label{subsec:Abel}

Given a merormophic form $\omega\in\mathcal{M}^h(\Sigma)$ of weight $h$, we know that its divisor $D=(\omega)$ must have degree $\deg{D}=(2g-2)h$. However, not every divisor of that degree can be expressed as the divisor of some meromorphic form. The question, then, is what are the necessary and sufficient conditions on $D$ such that $D=(\omega)$ for some form $\omega$? 

To start, a simpler question is to ask which divisors define a meromorphic function $f\in\mathcal{M}(\Sigma)$. To answer this question, consider the map $\widetilde{\mu}:\Sigma\to\mathbb{C}^g$ defined by

\begin{equation}\label{eq:abel-map}
\widetilde{\mu}(z)=\int_{p}^{z}\vec{\omega}\ ,
\end{equation}

\noindent where $p$ is some base-point and $\vec{\omega}=(\omega_1,\ldots,\omega_g)$ is the vector of canonically normalized holomorphic one-forms on $\Sigma$. However, note that the integration contour is not uniquely defined, and $\widetilde{\mu}$ is multivalued on $\mathbb{C}^g$. Indeed, one could wrap the contour around the $\alpha_{\mu}$ cycle $m_{\mu}$ times and around the $\beta_{\mu}$ cycle $n_{\mu}$ times and the integral would change by a factor

\begin{equation}
\widetilde{\mu}(z)\to\widetilde{\mu}(z)+\sum_{\mu}m_{\mu}\vec{e}_{\mu}+\sum_{\mu}n_{\mu}\left(\sum_{\nu}\Omega_{\mu\nu}\vec{e}_{\nu}\right)\ .
\end{equation}

\noindent That is, $\widetilde{\mu}$ is ambiguous up to the addition of an element of the lattice $\mathbb{Z}^g+\Omega\mathbb{Z}^g$. Thus, $\widetilde{\mu}(z)$ is unambiguous as an element of the quotient space $\mathbb{C}^g/(\mathbb{Z}^g+\Omega\mathbb{Z}^g)$, known as the \textit{Jacobian} of $\Sigma$, denoted by $\text{Jac}(\Sigma)$. We can therefore define a new function $\mu:\Sigma\to\text{Jac}(\Sigma)$ given by

\begin{equation}
\mu(z)=\int_{p}^{z}\vec{\omega}\quad(\text{mod }\mathbb{Z}^g+\Omega\mathbb{Z}^g)\ ,
\end{equation} 

\noindent which is independent of the integration path chosen. The function $\mu$ is known as the \textit{Abel-Jacobi map}. We can also extend the definition of $\mu$ to a linear homomorphism $\mu:\text{Div}(\Sigma)\to\text{Jac}(\Sigma)$. If $D=\sum_{i}n_ip_i$ is a divisor on $\Sigma$, then we can define

\begin{equation}
\mu(D)=\int_{p}^{D}\vec{\omega}=\sum_{i}n_i\int_{p}^{p_i}\vec{\omega_i}\quad(\text{mod }\mathbb{Z}^g+\Omega\mathbb{Z}^g)\ ,
\end{equation}

\noindent which maps the linear structure on $\text{Div}(\Sigma)$ to the linear structure on $\text{Jac}(\Sigma)$.

We can now provide a definitive answer to whether a meromorphic function $f$ with $(f)=D$ exists for any divisor $D$. Given a divisor $D$ with $\deg{D}=0$, there exists a unique meromorphic function (up to scalar multiplication) with $D=(f)$ if and only if $\mu(D)=0$. This is \textit{Abel's theorem}. A compact and elegant way to state Abel's theorem is that the short sequence of vector spaces

\begin{equation}
\mathcal{M}(\Sigma)\xrightarrow[\hspace{0.5cm}]{(\cdot)}\text{Div}_0(\Sigma)\xrightarrow[\hspace{0.5cm}]{\mu}\text{Jac}(\Sigma)
\end{equation}

\noindent is exact, where $\text{Div}_0(\Sigma)$ is the subspace of $\text{Div}(\Sigma)$ of divisors with degree zero, and $(\cdot):\mathcal{M}(\Sigma)\to\text{Div}_0(\Sigma)$ is the map taking a meromorphic function $f$ to its divisor $(f)$. Note also that Abel's theorem implies that $\mu(D)$ depends only on the equivalence class of the divisor $D$, since if $D'=D+(f)$ for any meromorphic function $f$, we have $\mu(D)=\mu(D')$.

Now we can answer the question for forms of generic weight $h$. Given two forms $\omega_1,\omega_2$ of degree $h$, note that their ratio $f(z)=\omega_1(z)/\omega_2(z)$ defines a meromorphic function,\footnote{This is true for integer $h$. For half-integer $h$, this is true if $\omega_1$ and $\omega_2$ are spinors on the same spin structure on $\Sigma$, so that the resulting function $f$ has no branch cuts.} and thus we have

\begin{equation}
\mu((\omega_1))-\mu((\omega_2))=\mu((\omega_1/\omega_2))=\mu((f))=0\ .
\end{equation}

\noindent That is, the divisors $(\omega_1)$ and $(\omega_2)$ have the same image under $\mu$. For $h=1$, recall that the divisor $(\omega)$ is equivalent to the canonical class $K$. The divisor of a meromorphic form of degree $h$ is equivalent to $hK$, and thus for any $\omega\in\mathcal{M}^{h}(\Sigma)$, we have

\begin{equation}
\mu((\omega))=h\mu(K)\ .
\end{equation}

\noindent Note that the definition of $\mu(K)$ depends on the choice of basepoint $p$ in \eqref{eq:abel-map}. Abel's theorem now extends to the construction of meromorphic forms, stating that a divisor $D$ with degree $\deg{D}=(2g-2)h$ is of the form $(\omega)$ if and only if $\mu(D)=h\mu(K)$. We will use this result extensively.

\subsection{Spinors and spin structures}\label{subsec:spinors}

So far, we have discussed meromorphic forms of weight $h$, with $h$ integer. However, in field theory, we also deal with conformal fields of weight $1/2$ (for example, the symplectic Bosons $\xi^{\pm},\eta^{\pm}$ of the main text). Furthermore, the prime form $\vartheta(x,y)$, which we discuss later, is defined as a form of weight $-1/2$ in both its arguments. Thus, we need to discuss forms of half-integer weight, as well as their divisors.

Let us start with meromorphic forms of weight $1/2$. Such a form can be written formally as $f(z)\sqrt{\mathrm{d}z}$. More formally, it is an object $\omega(z)$ which transforms like a conformal field of weight $1/2$ under coordinate transformations. The important subtlety is that we must choose a particular branch of the square root. This can be done unambiguously locally, but we must choose how to stitch together chosen branches globally in a consistent way. Such a stitching of branch choices is called a \textit{spin structure}, and on a surface of genus $g$ there are $2^{2g}$ such inequivalent spin structures, consisting of all the different choices of monodromy when transporting around the $2g$ cycles $\alpha_{\mu},\beta_{\mu}$.

Equivalently, we can think of a spin bundle on $\Sigma$ as a bundle $S$ such that $S^2=K$ -- i.e., a bundle whose sections square to meromorphic forms of degree 1. A choice of such a bundle is equivalent to a choice of spin structure, and there are again $2^{2g}$ such non-isomorphic choices. To such a spin bundle $S$, we associate a divisor class $\Delta$, such that $2\Delta$ is the canonical class on $\Sigma$. The degree of such a divisor class $\Delta$ is simply half the degree of the canonical class. That is,

\begin{equation}
\text{deg}(\Delta)=g-1\ .
\end{equation}

\noindent Put concretely, a meromorphic form of weight $1/2$ will always satisfy $Z(\omega)-P(\omega)=g-1$, where $Z$ counts the number of zeroes and $P$ counts the number of poles.

Finally we present one more way to think of spin structures in terms of the Jacobian lattice. A spin structure is represented by a divisor class $\Delta$ such that $2\Delta=K$, and thus under the Abel-Jacobi map we must have

\begin{equation}
2\mu(\Delta)\equiv 0\quad(\text{mod }\mathbb{Z}+\Omega\mathbb{Z})\ .
\end{equation}

\noindent Thus, we can write $\mu(\Delta)$ as

\begin{equation}
\mu(\Delta)=\frac{1}{2}\left(\vec{m}+\Omega\vec{n}\right)\ ,
\end{equation}

\noindent where $\vec{m},\vec{n}\in\mathbb{Z}^g$ are vectors of integers. Such an element of the lattice $\mathbb{Z}^g+\Omega\mathbb{Z}^g$ is called a \textit{half-period}. There are $2^{2g}$ such half-periods, one for each spin structure on $\Sigma$.

\subsection{Constructing functions and the prime form}\label{subsec:prime-form}

Given a divisor $D$ with $\deg{D}=0$ and $\mu(D)=0$, Abel's theorem tells us that there is a unique meromorphic function $f:\Sigma\to\mathbb{C}$ with divisor $D$, up to scalar multiplication. It is then natural to ask if there is a generic way to construct such a function out of elementary components. On the sphere, we know this is possible. If $D=\sum_in_iz_i$ of points $z_i$ with multiplicities $n_i$ such that $\sum_{i}n_i=0$, the expression

\begin{equation}
f(z)=\prod_{i}(z-z_i)^{n_i}
\end{equation}

\noindent defines such a meromorphic function, built from the fundamental building block $(x-y)$.

One could then ask whether a function analogous to $(x-y)$ exists on a Riemann surface of genus $g$. The answer is, in fact, yes, and it is given by the so-called ``prime form'' $\vartheta(x,y)$. The prime form is constructed using objects called theta functions, but we will not need the explicit form of its construction here (the interested reader is encouraged to consult Chapter 3 of \cite{eynard2018lectures}). For our purposes, it suffices to know that the prime form satisfies the following properties:\footnote{Technically, the construction of the prime form requires one to pick a spin structure on $\Sigma$. We will ignore this subtlety, since it does not enter in our analysis. See \cite{eynard2018lectures} for more details.}

\begin{itemize}

	\item $\vartheta(x,y)$ is a meromorphic form of weight $-\frac{1}{2}$ in both $x$ and $y$.

	\item $\vartheta(x,y)$ has a simple zero only when $x=y$.

	\item $\vartheta(x,y)=-\vartheta(y,x)$.

	\item $\vartheta(x,y)$ is periodic around the $\alpha$-cycles, but is only quasi-periodic around the $\beta$-cycles, picking up the monodromy

	\begin{equation}
	\vartheta(x+\beta_\mu,y)=\exp\left(-i\pi\Omega_{\mu\mu}-2\pi i\int_{x}^{y}\omega_{\mu}\right)\vartheta(x,y)\ .
	\end{equation}

\end{itemize}

\noindent Assuming that such an object with these properties exists, we claim that we can use it to construct meromorphic forms from their divisors. Given a divisor $D=\sum_{i}n_iz_i$ with $\mu(D)=0$ and $\deg{D}=\sum_{i}n_i=0$, the expression

\begin{equation}\label{eq:function-from-prime-form}
f(z)=\prod_{i}\vartheta(z,z_i)^{n_i}
\end{equation}

\noindent is a globally-defined meromorphic function on $\Sigma$ with divisor $D$. It is clear that the function $f(z)$ defined above has divisor $D$. Furthermore, the function $f(z)$ is well-defined on the surface $\Sigma$ (i.e. it is periodic around the cycles of $\Sigma$). Indeed, we have

\begin{equation}
\begin{split}
f(z+\beta_{\mu})&=\prod_{i}\vartheta(z+\beta_{\mu},z_i)^{n_i}\\
&=\prod_{i}\left(\exp\left(-i\pi \Omega_{\mu\mu}-2\pi i\int_{z}^{z_i}\omega_{\mu}\right)^{n_i}\right)f(z)\\
&=\exp\left(-i\pi\Omega_{\mu\mu}\sum_{i}n_i-2\pi i\sum_{i}n_i\int_{z}^{z_i}\omega_{\mu}\right)f(z)\\
&=\exp\left(-i\pi\deg{D}\,\Omega_{\mu\mu}-2\pi i\mu(D)\right)f(z)\ ,
\end{split}
\end{equation}

\noindent and the monodromy vanishes if $\deg{D}=0$ and $\mu(D)=0$. Periodicity around the $\alpha$ cycles is trivial since $\vartheta(x,y)$ is periodic around the $\alpha$ cycles. Thus, \eqref{eq:function-from-prime-form} is a globally defined meromorphic function with the desired divisor. By Abel's theorem, this function is unique.

\section{\boldmath Localization of the \texorpdfstring{$W$}{W} fields}\label{sec:u-dependence}

In this appendix we discuss the dependence of the correlation functions on the coordinates $u_{\alpha}$. In particular, we note that these coordinates also localize to a codimension-$g$ subspace. We argue that this localization is natural from the point of view of the gauging of the $\widehat{\mathfrak{u}(1)}$ subalgebra in the symplectic Boson theory\footnote{Thank you to Lorenz Eberhardt, whose helpful conversations illuminated this particular point.}.

The correlation functions

\begin{equation}\label{eq:correlators-u}
\Bigl\langle\prod_{\alpha=1}^{n+2g-2}W(u_{\alpha})\prod_{i=1}^{n}V_{m_i,j_i}^{w_i}(x_i;z_i)\Bigr\rangle
\end{equation}

\noindent in the free field construction of strings on $\rm{AdS}_3$ carry a nontrivial dependence on the coordinates $u_{\alpha}$. However, under the projection relating $\mathfrak{sl}(2,\mathbb{R})_1\oplus\widehat{\mathfrak{u}(1)}$ to $\mathfrak{sl}(2,\mathbb{R})_1$, the field $W$ maps to the identity, and thus the coordinates $u_{\alpha}$ are unphysical with respect to string theory on $\rm{AdS}_3$. In order to relate \eqref{eq:correlators-u} to physical correlation functions on $\rm{AdS}_3$ string theory, it is important to determine the $u_{\alpha}$-dependence of \eqref{eq:correlators-u}. On the sphere, this dependence is simply fixed by the $\mathfrak{u}(1)$ Knizhnik-Zamolodchikov equations \cite{Knizhnik:1984nr}. However, at higher genus, the dependence on the $u_{\alpha}$ coordinates is much harder to pin down, and will not be done here. Moreover, unlike for correlation functions on the sphere, the coordinates $u_{\alpha}$ obey $g$ extra localization conditions, which we demonstrate below. The following discussion is rather informal, and does not significantly affect the main text. More details on this particular point appear in \cite{Eberhardt:2021}.

In the main text, we defined the spinor forms $\omega^{\pm}(z)$ on $\Sigma$ by

\begin{equation}
\omega^{\pm}(z)=\Big\langle\xi^{\pm}(z)\prod_{\alpha=1}^{n+2g-2}W(u_{\alpha})\prod_{i=1}^{n}V_{m_i,j_i}^{w_i}(x_i;z_i)\Big\rangle\ .
\end{equation}

\noindent We found that the divisors of $\omega^{\pm}$ can be written in the form

\begin{equation}\label{eq:omega-divisor}
(\omega^{\pm})=-\sum_{i=1}^{n}\frac{w_i+1}{2}z_i+\sum_{\alpha=1}^{n+2g-2}u_{\alpha}+\Delta^{\pm}
\end{equation}

\noindent for some positive divisors $\Delta^{\pm}$. We then, using Abel's theorem as well as the OPEs \eqref{eq:regular-OPE}, showed that the forms $\omega^{\pm}(z)$ satisfied all of the algebraic relations needed to construct a covering map $\Gamma$ with divisor $(\Gamma)=\Delta^--\Delta^+$. This left us with a set of $2N+n+3g-2$ constraints in $2N+1$ unknowns, effectively demonstrating that the correlation functions localized to points in the moduli space for which $\Sigma$ is a branched covering of the CFT sphere.

Now, let us take the insertion points $z_i$, the moduli of $\Sigma$, and the divisors $\Delta^{\pm}$ to be fixed, determined by some covering map $\Gamma$. Then applying the Abel-Jacobi map $\mu$ to \eqref{eq:omega-divisor} gives us

\begin{equation}
\mu((\omega^{\pm}))=-\sum_{i=1}^{n}\frac{w_i+1}{2}\mu(z_i)+\sum_{\alpha=1}^{n+2g-2}\mu(u_{\alpha})+\mu(\Delta^{\pm})\ .
\end{equation}

\noindent In the main text, we used the fact that $\mu((\omega^+))=\mu((\omega^-))$ to show that $\mu(\Delta^+)=\mu(\Delta^-)$. However, since $\omega^{\pm}$ are spinor forms, we must have $\mu((\omega^{\pm}))=\mu(\Delta)$, where $\Delta$ is the divisor class of our chosen spin structure. Thus, we have

\begin{equation}\label{eq:u-constraints}
\sum_{\alpha=1}^{n+2g-2}\mu(u_{\alpha})=\sum_{i=1}^{n}\frac{w_i+1}{2}\mu(z_i)-\mu(\Delta^{\pm})+\mu(\Delta)\ .
\end{equation}

\noindent The right-hand-side of this equation is expressed purely in terms of fixed quantities, and thus defines a set of constraints on the coefficients $u_{\alpha}$. In particular, we see that the coefficients $u_{\alpha}$ must be localized to a codimension $g$ locus for which \eqref{eq:u-constraints} is satisfied. That is, the correlation functions \eqref{eq:generic-correlator} do not just satisfy a localization principle for the moduli of $\Sigma\setminus\{z_1,\ldots,z_n\}$, but also a localization property for the ``unphysical'' coordinates $u_{\alpha}$.

To understand this localization property a little better, let us recall that in order to retrieve correlation functions which describe scattering in tensionless $\rm{AdS}_3$ string theory, we must gauge out the $\widehat{\mathfrak{u}(1)}$ subsector generated by the current $U(z)$. One convenient way to perform this gauging is to introduce ghost fields $b$ and $c$ with conformal weights $h(b)=1$ and $h(c)=0$, and introduce a BRST operator $Q\propto \oint (c\,U)(z)$ \cite{Karabali:1989dk}. Then gauging the $\widehat{\mathfrak{u}(1)}$ algebra is implemented by the standard BRST quantization with respect to the charge $Q$. At the end of the day, this procedure will result in a correlation function which is expressed as an integral over the zero modes of $b(z)$, just as in the BRST quantization of the Bosonic string. Each zero mode of $b$ corresponds to a holomorphic differential on $\Sigma$, and so this is a $g$-dimensional integral. Equivalently, we integrate over the moduli space of all flat $\mathfrak{u}(1)$ bundles on $\Sigma$, which is isomorphic to the Jacobian $\text{Jac}(\Sigma)\cong\mathbb{C}^g/\left(\mathbb{Z}+\Omega\mathbb{Z}\right)$.

Now, the ``physical'' correlation function is obtained by the correlators on the worldsheet by some appropriate integration over $\mathcal{M}_{g,n}$ and $\text{Jac}(\Sigma)$. That is, the physical quantities of interest, which are matched to correlators in the dual CFT, are (schematically) given by

\begin{equation}\label{eq:quotiented-correlator}
\int_{\mathcal{M}_{g,n}\times\text{Jac}(\Sigma)}\mathrm{d}\mu\,\Big\langle\prod_{\alpha=1}^{n+2g-2}W(u_{\alpha})\prod_{i=1}^{n}V_{m_i,j_i}^{w_i}(x_i;z_i)\Big\rangle
\end{equation}

\noindent for some appropriate measure $\mathrm{d}\mu$. The condition \eqref{eq:u-constraints} then suggests that these correlators localize to one point on the Jacobian, which, along with the localization on $\mathcal{M}_{g,n}$ demonstrated in the main text, suggest that only one point in this integral contributes to the final result.

Finally, we note that \eqref{eq:u-constraints} depends explicitly on the spin structure $\Delta$ we choose on $\Sigma$. Since any spin bundle $S\to\Sigma$ can be obtained from some fixed spin bundle by tensoring with a $\mathbb{Z}_2\subset\mathbb{C}^{\times}$ bundle, the sum over all spin bundles can be naturally included in the integral over all nontrivial $\mathfrak{u}(1)$ bundles over $\Sigma$ (see Section 4 of \cite{Eberhardt:2021}).

\section{\boldmath The \texorpdfstring{$g/2$}{g/2}-form}\label{appendix:sigma}

In Section \ref{sec:Ward-to-covering} we used the existence of a quasiperiodic meromorphic $g/2$-form, which we called $\sigma(z)$, to construct correlation functions with $\xi^{\pm}$ insertions. In this section, we review the construction of such a form, demonstrating that it exists, and furthermore showing that it is unique up to a constant multiple.

Let us first recall the properties of $\sigma$:

\begin{itemize}

	\item $\sigma$ transforms as a meromorphic $g/2$-form on $\Sigma$.

	\item $\sigma$ has no zeroes or poles.

	\item $\sigma$ is periodic around the $\alpha$-cylces of $\Sigma$ but is only quasiperiodic around the $\beta$-cycles, along which it picks up a nontrivial monodromy,

	\begin{equation}
	\sigma(z+\beta_{\mu})=\exp\left(-i\pi(g-1)\Omega_{\mu\mu}-2\pi i\int_{(g-1)z}^{\Delta}\omega_{\mu}\right)\sigma(z)\ .
	\end{equation}

\end{itemize}

\noindent Once we have constructed $\sigma(z)$, its uniqueness is fairly straightforward to verify. Let us assume that two such $g/2$-forms exist, and call them $\sigma(z)$ and $\widetilde{\sigma}(z)$. Both $\sigma$ and $\widetilde{\sigma}$ are holomorphic functions on the covering space $\widetilde{\Sigma}_g$ and obtain the same monodromy around the $\beta$-cycles of $\Sigma$. Furthermore, both $\sigma$ and $\widetilde{\sigma}$ have no zeroes. Thus, the function

\begin{equation}
f(z):=\frac{\sigma(z)}{\widetilde{\sigma}(z)}
\end{equation}

\noindent is holomorphic and globally defined on $\Sigma$. Since the only holomorphic functions on compact Riemann surfaces are constants, we are left to conclude that $\sigma$ and $\widetilde{\sigma}$ are related by a constant scalar multiple. Thus, once we have constructed any $g/2$-form which satisfies these properties, we are done.

In order to construct $\sigma$, consider any meromorphic 1-form $\omega\in\mathcal{M}^1(\Sigma)$ with divisor

\begin{equation}
D=\sum_{a}Q_a-\sum_{a}P_a\ .
\end{equation}

\noindent Then let us define

\begin{equation}
\sigma(z)=\sqrt{\prod_{a}\vartheta(z,P_a)\,\prod_{a}\vartheta(z,Q_a)^{-1}\omega(z)}\ .
\end{equation}

\noindent Since $\deg{D}=2(g-1)$, this defines a meromorphic form of weight $g/2$. Furthermore, the argument of the square root has no zeroes and no poles, and thus taking the square root is unambiguous up to a sign. Finally, from the quasiperiodicity of the prime form $\vartheta(x,y)$, one easily shows that $\sigma(z)$ picks up the desired monodromy around the $\beta$-cycles, since the image of $D$ under the Abel-Jacobi map is the canonical class of $\Sigma$.

\bibliography{draft2}

\providecommand{\href}[2]{#2}\begingroup\raggedright\begin{thebibliography}{10}

\bibitem{Dei:2020}
A.~Dei, M.~R. Gaberdiel, R.~Gopakumar, and B.~Knighton, ``{Free field
  world-sheet correlators for ${\rm AdS}_3$},''
  \href{http://dx.doi.org/10.1007/JHEP02(2021)081}{{\em JHEP} {\bfseries 02}
  (2021) 081}, \href{http://arxiv.org/abs/2009.11306}{{\ttfamily
  arXiv:2009.11306 [hep-th]}}.

\bibitem{Maldacena_1999}
J.~Maldacena, ``{The Large $N$ Limit of Superconformal Field Theories and
  Supergravity},'' \href{http://dx.doi.org/10.1023/a:1026654312961}{{\em
  International Journal of Theoretical Physics} no.~4, (1999) 1113–1133},
  \href{http://arxiv.org/abs/hep-th/9711200}{{\ttfamily arXiv:hep-th/9711200}}.

\bibitem{Gaberdiel_2018}
M.~R. Gaberdiel and R.~Gopakumar, ``{Tensionless string spectra on
  AdS$_{3}$},'' \href{http://dx.doi.org/10.1007/JHEP05(2018)085}{{\em JHEP}
  {\bfseries 05} (2018) 085}, \href{http://arxiv.org/abs/1803.04423}{{\ttfamily
  arXiv:1803.04423 [hep-th]}}.

\bibitem{Eberhardt:2018}
L.~Eberhardt, M.~R. Gaberdiel, and R.~Gopakumar, ``{The Worldsheet Dual of the
  Symmetric Product CFT},''
  \href{http://dx.doi.org/10.1007/JHEP04(2019)103}{{\em JHEP} {\bfseries 04}
  (2019) 103}, \href{http://arxiv.org/abs/1812.01007}{{\ttfamily
  arXiv:1812.01007 [hep-th]}}.

\bibitem{Eberhardt:2019}
L.~Eberhardt, M.~R. Gaberdiel, and R.~Gopakumar, ``{Deriving the
  AdS$_{3}$/CFT$_{2}$ correspondence},''
  \href{http://dx.doi.org/10.1007/JHEP02(2020)136}{{\em JHEP} {\bfseries 02}
  (2020) 136}, \href{http://arxiv.org/abs/1911.00378}{{\ttfamily
  arXiv:1911.00378 [hep-th]}}.

\bibitem{Eberhardt:2020}
L.~Eberhardt, ``{AdS$_{3}$/CFT$_{2}$ at higher genus},''
  \href{http://dx.doi.org/10.1007/JHEP05(2020)150}{{\em JHEP} {\bfseries 05}
  (2020) 150}, \href{http://arxiv.org/abs/2002.11729}{{\ttfamily
  arXiv:2002.11729 [hep-th]}}.

\bibitem{Hikida_2020}
Y.~Hikida and T.~Liu, ``{Correlation functions of symmetric orbifold from
  AdS$_{3}$ string theory},''
  \href{http://dx.doi.org/10.1007/JHEP09(2020)157}{{\em JHEP} {\bfseries 09}
  (2020) 157}, \href{http://arxiv.org/abs/2005.12511}{{\ttfamily
  arXiv:2005.12511 [hep-th]}}.

\bibitem{Sfondrini:2020ovj}
A.~Sfondrini, ``{Long Strings and Symmetric Product Orbifold from the AdS$_3$
  Bethe Equations},'' \href{http://arxiv.org/abs/2010.02782}{{\ttfamily
  arXiv:2010.02782 [hep-th]}}.

\bibitem{Gaberdiel:2020ycd}
M.~R. Gaberdiel, R.~Gopakumar, B.~Knighton, and P.~Maity, ``{From Symmetric
  Product CFTs to ${\rm AdS}_3$},''
  \href{http://arxiv.org/abs/2011.10038}{{\ttfamily arXiv:2011.10038
  [hep-th]}}.

\bibitem{Hamidi:1986vh}
S.~Hamidi and C.~Vafa, ``{Interactions on Orbifolds},''
  \href{http://dx.doi.org/10.1016/0550-3213(87)90006-X}{{\em Nucl. Phys. B}
  {\bfseries 279} (1987) 465--513}.

\bibitem{Lunin_2001}
O.~Lunin and S.~D. Mathur, ``{Correlation Functions for $M_N/S_N$ Orbifolds},''
  \href{http://dx.doi.org/10.1007/s002200100431}{{\em Communications in
  Mathematical Physics} {\bfseries 219} no.~2, (May, 2001) 399–442}.

\bibitem{Lunin_2002}
O.~Lunin and S.~D. Mathur, ``{Three-Point Functions for $M_N/S_N$ Orbifolds
  with $\mathcal{N} = 4$ Supersymmetry},''
  \href{http://dx.doi.org/10.1007/s002200200638}{{\em Communications in
  Mathematical Physics} {\bfseries 227} no.~2, (May, 2002) 385–419},
  \href{http://arxiv.org/abs/hep-th/0103169}{{\ttfamily arXiv:hep-th/0103169}}.

\bibitem{Pakman_2009}
A.~Pakman, L.~Rastelli, and S.~S. Razamat, ``{Diagrams for Symmetric Product
  Orbifolds},'' \href{http://dx.doi.org/10.1088/1126-6708/2009/10/034}{{\em
  JHEP} {\bfseries 10} (2009) 034},
  \href{http://arxiv.org/abs/0905.3448}{{\ttfamily arXiv:0905.3448 [hep-th]}}.

\bibitem{Berkovits_1999}
N.~Berkovits, C.~Vafa, and E.~Witten, ``{Conformal field theory of AdS
  background with Ramond-Ramond flux},''
  \href{http://dx.doi.org/10.1088/1126-6708/1999/03/018}{{\em JHEP} {\bfseries
  03} (1999) 018}, \href{http://arxiv.org/abs/hep-th/9902098}{{\ttfamily
  arXiv:hep-th/9902098}}.

\bibitem{Maldacena-Ooguri-1}
J.~Maldacena and H.~Ooguri, ``{Strings in $AdS_3$ and the $\text{SL}(2,R)$ WZW
  model. I: The spectrum},'' \href{http://dx.doi.org/10.1063/1.1377273}{{\em
  Journal of Mathematical Physics} {\bfseries 42} no.~7, (Jul, 2001)
  2929–2960}, \href{http://arxiv.org/abs/hep-th/0001053}{{\ttfamily
  arXiv:hep-th/0001053}}.

\bibitem{Maldacena-Ooguri-2}
J.~Maldacena, H.~Ooguri, and J.~Son, ``{Strings in $AdS_3$ and the
  $\text{SL}(2,R)$ WZW model. II: Euclidean black hole},''
  \href{http://dx.doi.org/10.1063/1.1377039}{{\em Journal of Mathematical
  Physics} {\bfseries 42} no.~7, (Jul, 2001) 2961–2977},
  \href{http://arxiv.org/abs/hep-th/0005183}{{\ttfamily arXiv:hep-th/0005183}}.

\bibitem{Maldacena-Ooguri-3}
J.~Maldacena and H.~Ooguri, ``{Strings in $AdS_3$ and the $\text{SL}(2,R)$ WZW
  model. III. Correlation functions},''
  \href{http://dx.doi.org/10.1103/physrevd.65.106006}{{\em Physical Review D}
  {\bfseries 65} no.~10, (May, 2002) },
  \href{http://arxiv.org/abs/hep-th/0111180}{{\ttfamily arXiv:hep-th/0111180}}.

\bibitem{Ferreira:2017}
K.~Ferreira, M.~R. Gaberdiel, and J.~I. Jottar, ``{Higher spins on AdS$_{3}$
  from the worldsheet},'' \href{http://dx.doi.org/10.1007/JHEP07(2017)131}{{\em
  JHEP} {\bfseries 07} (2017) 131},
  \href{http://arxiv.org/abs/1704.08667}{{\ttfamily arXiv:1704.08667
  [hep-th]}}.

\bibitem{Goddard:1987}
P.~Goddard, D.~I. Olive, and G.~Waterson, ``{Superalgebras, Symplectic Bosons
  and the Sugawara Construction},''
  \href{http://dx.doi.org/10.1007/BF01225374}{{\em Commun. Math. Phys.}
  {\bfseries 112} (1987) 591}.

\bibitem{DiFrancesco}
P.~Francesco, P.~Mathieu, and D.~S\'en\'echal, {\em Conformal field Theory}.
\newblock New York: Springer-Verlag Press, 1997.

\bibitem{Dei:2019}
A.~Dei and L.~Eberhardt, ``{Correlators of the symmetric product orbifold},''
  \href{http://dx.doi.org/10.1007/JHEP01(2020)108}{{\em JHEP} {\bfseries 01}
  (2020) 108}, \href{http://arxiv.org/abs/1911.08485}{{\ttfamily
  arXiv:1911.08485 [hep-th]}}.

\bibitem{Berkovits:1994vy}
N.~Berkovits and C.~Vafa, ``{$\mathcal{N}=4$ topological strings},''
  \href{http://dx.doi.org/10.1016/0550-3213(94)00419-F}{{\em Nucl. Phys. B}
  {\bfseries 433} (1995) 123--180},
  \href{http://arxiv.org/abs/hep-th/9407190}{{\ttfamily arXiv:hep-th/9407190}}.

\bibitem{Friedan:1985ge}
D.~Friedan, E.~J. Martinec, and S.~H. Shenker, ``{Conformal Invariance,
  Supersymmetry and String Theory},''
  \href{http://dx.doi.org/10.1016/0550-3213(86)90356-1}{{\em Nucl. Phys. B}
  {\bfseries 271} (1986) 93--165}.

\bibitem{Berkovits_2004}
N.~Berkovits, ``{Alternative String Theory in Twistor Space for $\mathcal{N}=4$
  Super-Yang-Mills Theory},''
  \href{http://dx.doi.org/10.1103/physrevlett.93.011601}{{\em Physical Review
  Letters} {\bfseries 93} no.~1, (Jun, 2004) },
  \href{http://arxiv.org/abs/hep-th/0402045}{{\ttfamily arXiv:hep-th/0402045}}.

\bibitem{Verlinde:1990ku}
E.~P. Verlinde and H.~L. Verlinde, ``{A Solution of Two-dimensional Topological
  Quantum Gravity},''
  \href{http://dx.doi.org/10.1016/0550-3213(91)90200-H}{{\em Nucl. Phys. B}
  {\bfseries 348} (1991) 457--489}.

\bibitem{Belavin:2006ex}
A.~Belavin and A.~Zamolodchikov, ``{Integrals over moduli spaces, ground ring,
  and four-point function in minimal Liouville gravity},''
  \href{http://dx.doi.org/10.1007/s11232-006-0075-8}{{\em Theor. Math. Phys.}
  {\bfseries 147} (2006) 729--754}.

\bibitem{Aharony:2006th}
O.~Aharony, Z.~Komargodski, and S.~S. Razamat, ``{On the worldsheet theories of
  strings dual to free large N gauge theories},''
  \href{http://dx.doi.org/10.1088/1126-6708/2006/05/016}{{\em JHEP} {\bfseries
  05} (2006) 016}, \href{http://arxiv.org/abs/hep-th/0602226}{{\ttfamily
  arXiv:hep-th/0602226}}.

\bibitem{Aharony:2007fs}
O.~Aharony, J.~R. David, R.~Gopakumar, Z.~Komargodski, and S.~S. Razamat,
  ``{Comments on worldsheet theories dual to free large N gauge theories},''
  \href{http://dx.doi.org/10.1103/PhysRevD.75.106006}{{\em Phys. Rev. D}
  {\bfseries 75} (2007) 106006},
  \href{http://arxiv.org/abs/hep-th/0703141}{{\ttfamily arXiv:hep-th/0703141}}.

\bibitem{Razamat:2008zr}
S.~S. Razamat, ``{On a worldsheet dual of the Gaussian matrix model},''
  \href{http://dx.doi.org/10.1088/1126-6708/2008/07/026}{{\em JHEP} {\bfseries
  07} (2008) 026}, \href{http://arxiv.org/abs/0803.2681}{{\ttfamily
  arXiv:0803.2681 [hep-th]}}.

\bibitem{Verlinde:1987}
E.~P. Verlinde and H.~L. Verlinde, ``{Chiral Bosonization, Determinants and the
  String Partition Function},''
  \href{http://dx.doi.org/10.1016/0550-3213(87)90219-7}{{\em Nucl. Phys. B}
  {\bfseries 288} (1987) 357}.

\bibitem{eberhardt2020partition}
L.~Eberhardt, ``Partition functions of the tensionless string,''
  \href{http://arxiv.org/abs/2008.07533}{{\ttfamily arXiv:2008.07533
  [hep-th]}}.

\bibitem{Bachas:2002-perm}
C.~Bachas, J.~d. Boer, R.~Dijkgraaf, and H.~Ooguri, ``Permeable conformal walls
  and holography,'' \href{http://dx.doi.org/10.1088/1126-6708/2002/06/027}{{\em
  Journal of High Energy Physics} {\bfseries 2002} no.~06, (Jun, 2002)
  027–027}, \href{http://arxiv.org/abs/hep-th/0111210}{{\ttfamily
  arXiv:hep-th/0111210}}.

\bibitem{Bachas:2002-asymp}
C.~Bachas, ``{Asymptotic symmetries of AdS$_2$ branes},'' in {\em {Meeting on
  Strings and Gravity: Tying the Forces Together}}, pp.~9--17.
\newblock 2001.
\newblock \href{http://arxiv.org/abs/hep-th/0205115}{{\ttfamily
  arXiv:hep-th/0205115}}.

\bibitem{Eberhardt:2019niq}
L.~Eberhardt and M.~R. Gaberdiel, ``{Strings on $\text{AdS}_3 \times \text{S}^3
  \times \text{S}^3 \times \text{S}^1$},''
  \href{http://dx.doi.org/10.1007/JHEP06(2019)035}{{\em JHEP} {\bfseries 06}
  (2019) 035}, \href{http://arxiv.org/abs/1904.01585}{{\ttfamily
  arXiv:1904.01585 [hep-th]}}.

\bibitem{Eberhardt:2017fsi}
L.~Eberhardt, M.~R. Gaberdiel, R.~Gopakumar, and W.~Li, ``{BPS spectrum on
  $\rm{AdS}_3\times\rm{S}^3\times\rm{S}^3\times\rm{S}^1$},''
  \href{http://dx.doi.org/10.1007/JHEP03(2017)124}{{\em JHEP} {\bfseries 03}
  (2017) 124}, \href{http://arxiv.org/abs/1701.03552}{{\ttfamily
  arXiv:1701.03552 [hep-th]}}.

\bibitem{Eberhardt:2017pty}
L.~Eberhardt, M.~R. Gaberdiel, and W.~Li, ``{A holographic dual for string
  theory on $\rm{AdS}_3\times\rm{S}^3\times\rm{S}^3\times\rm{S}^1$},''
  \href{http://dx.doi.org/10.1007/JHEP08(2017)111}{{\em JHEP} {\bfseries 08}
  (2017) 111}, \href{http://arxiv.org/abs/1707.02705}{{\ttfamily
  arXiv:1707.02705 [hep-th]}}.

\bibitem{donaldson_2011}
S.~K. Donaldson, {\em Riemann surfaces}.
\newblock Oxford University Press, 2011.

\bibitem{eynard2018lectures}
B.~Eynard, ``{Lectures notes on compact Riemann surfaces},''
  \href{http://arxiv.org/abs/1805.06405}{{\ttfamily arXiv:1805.06405
  [math-ph]}}.

\bibitem{Knizhnik:1984nr}
V.~Knizhnik and A.~Zamolodchikov, ``{Current Algebra and Wess-Zumino Model in
  Two-Dimensions},'' \href{http://dx.doi.org/10.1016/0550-3213(84)90374-2}{{\em
  Nucl. Phys. B} {\bfseries 247} (1984) 83--103}.

\bibitem{Eberhardt:2021}
L.~Eberhardt, ``{Summing over Geometries in String Theory},''
  \href{http://arxiv.org/abs/2102.12355}{{\ttfamily arXiv:2102.12355
  [hep-th]}}.

\bibitem{Karabali:1989dk}
D.~Karabali and H.~J. Schnitzer, ``{BRST Quantization of the Gauged WZW Action
  and Coset Conformal Field Theories},''
  \href{http://dx.doi.org/10.1016/0550-3213(90)90075-O}{{\em Nucl. Phys. B}
  {\bfseries 329} (1990) 649--666}.

\end{thebibliography}\endgroup
\bibliographystyle{utphys.bst}

\end{document}